\documentclass{osa-article}
\journal{oe}

\usepackage{amsmath}

\DeclareMathOperator*{\argmin}{arg\,min}

\begin{document}

\title{Diffraction tomography with a deep image prior}

\author{Kevin C. Zhou\authormark{1,2} and Roarke Horstmeyer\authormark{1,3}}

\address{\authormark{1}Department of Biomedical Engineering, Duke University, Durham, NC 27708\\
\authormark{2}kevin.zhou@duke.edu\\
\authormark{3}roarke.w.horstmeyer@duke.edu}

\begin{abstract*}
We present a tomographic imaging technique, termed Deep Prior Diffraction Tomography (DP-DT), to reconstruct the 3D refractive index (RI) of thick biological samples at high resolution from a sequence of low-resolution images collected under angularly varying illumination. DP-DT processes the multi-angle data using a phase retrieval algorithm that is extended by a deep image prior (DIP), which reparameterizes the 3D sample reconstruction with an untrained, deep generative 3D convolutional neural network (CNN). We show that DP-DT effectively addresses the missing cone problem, which otherwise degrades the resolution and quality of standard 3D reconstruction algorithms. As DP-DT does not require pre-captured data or pre-training, it is not biased towards any particular dataset. Hence, it is a general technique that can be applied to a wide variety of 3D samples, including scenarios in which large datasets for supervised training would be infeasible or expensive. We applied DP-DT to obtain 3D RI maps of bead phantoms and complex biological specimens, both in simulation and experiment, and show that DP-DT produces higher-quality results than standard regularization techniques. We further demonstrate the generality of DP-DT, using two different scattering models, the first Born and multi-slice models. Our results point to the potential benefits of DP-DT for other 3D imaging modalities, including X-ray computed tomography, magnetic resonance imaging, and electron microscopy.
\end{abstract*}

\section{Introduction}
There are a variety of microscopes that can obtain high-resolution images in three dimensions, including scanning confocal microscopes, two-photon microscopes, and light sheet microscopes, to name a few \cite{mertz2019strategies}. Most of these methods are geared towards incoherent fluorescent imaging, and cannot produce a quantitative estimation of the 3D refractive index (RI) distributions of thick biological samples. Quantitative RI is useful for a number of reasons - it does not require labeling with fluorescent proteins or dyes, it can directly yield useful measures of cell mass and protein concentration, and it can provide useful information about how light is scattered in thick samples for subsequent imaging system correction, for example \cite{park2018quantitative}.

Currently, the primary technique for quantitative 3D RI measurement is diffraction tomography (DT) \cite{Wolf_1969, Sung_2009, Fiolka_2009, Lauer_2002, chowdhury2017refractive}. The first implementations of DT relied on holography to measure the complex field scattered from an object under illumination from a variety of angles. Since DT is a phase-sensitive technique, it requires a highly coherent beam with interferometric stability and some sort of angular scanning mechanism to steer the incident beam through a range of angles. These requirements make its practical implementation relatively complicated and challenging. 
Several recent techniques have demonstrated DT without a reference beam, instead using intensity-only images and a suitable phase retrieval algorithm for 3D sample reconstruction \cite{Horstmeyer_2016, chowdhury2019high, tian20153d, li2019high, ling2018high, pham2018versatile, jiang2016inverse}. These methods, which effectively extend Fourier ptychography (FP) techniques into the third dimension~\cite{Zheng_2013}, remove the need for a highly coherent beam and interferometric stability. Instead, they rely on a programmable LED array to provide angularly-varying illumination, which leads to a simple and compact device that requires no moving parts~\cite{Konda2019}. While a direct extension of FP into 3D relies on the first Born approximation, other related methods have also used the multi-slice (MS) model (also known as the beam propagation method, BPM) which can incorporate the effects of multiple-scattering \cite{kamilov2015learning, kamilov2016optical, tian20153d, chowdhury2019high, jiang2016inverse, pham2018versatile}.     

Since all of the above techniques illuminate a stationary sample from a finite angular extent, and use a single objective lens to collect the scattered light, they all suffer from the missing cone problem (also referred to as the missing wedge) \cite{Tam_1981}. In both phase-sensitive and intensity-only diffraction tomography, the missing cone manifests itself as a bandlimited transfer function that is zero within a cone surrounding the $k_z$ axis. This limited transfer function produces axial artifacts and underestimates the RI, thus presenting challenges to the accurate 3D reconstruction of thick samples at high resolution.

\begin{figure}[t]
\begin{center}
\includegraphics[width=0.8\columnwidth]{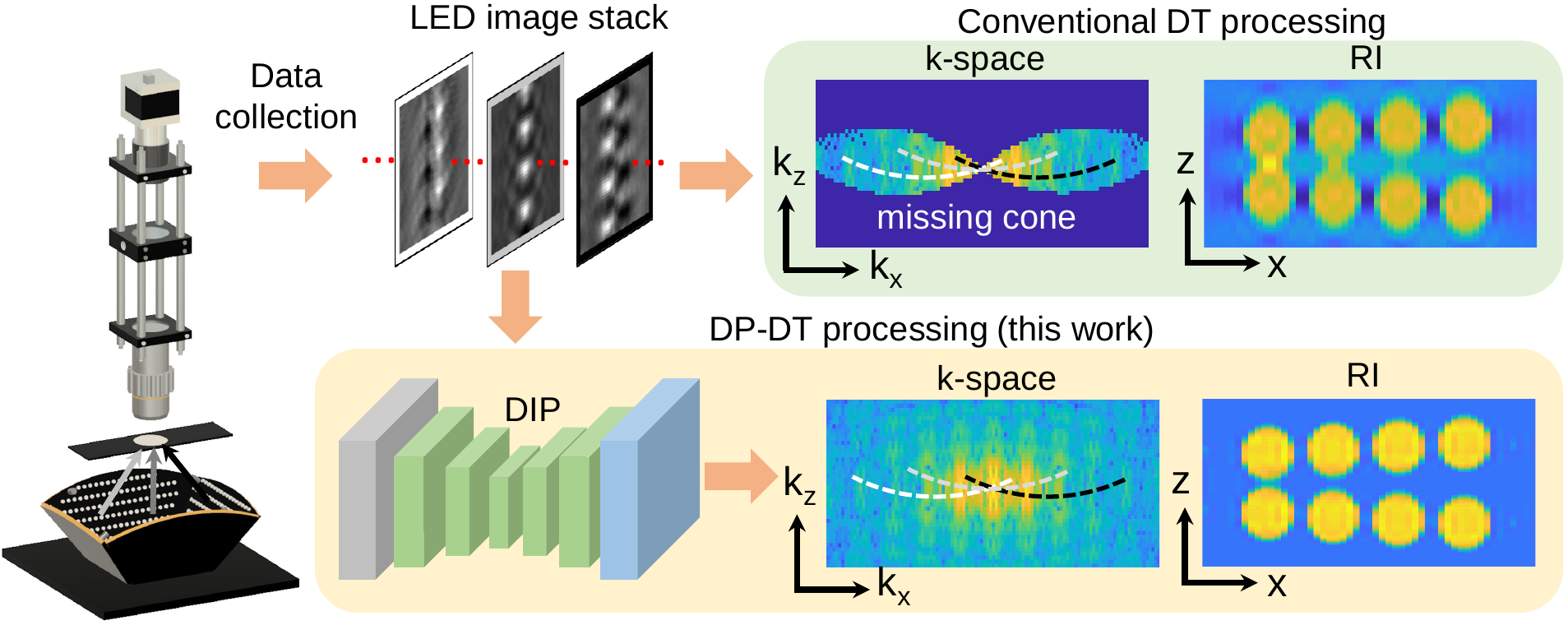}
\caption{Summary of Deep Prior Diffraction Tomography (DP-DT). Microscope captures variably-illuminated image $k$-stack. Standard tomographic reconstruction methods suffer from artifacts caused by a missing cone in $k$-space. In this work, we propose the use of a deep image prior to help account for missing cone artifacts to improve 3D image reconstruction.  
\label{teaser}%
}
\end{center}
\end{figure}

As we will show, the missing cone problem becomes significantly worse when using low numerical aperture (NA) objective lenses, and thus high-NA objectives are desirable for DT. On the other hand, low-NA objective lenses typically exhibit larger space-bandwidth products (SBPs) than high-NA lenses, since they can image over larger fields of view (FOVs). They also exhibit greater depths of field (and therefore high signal-to-noise ratio (SNR) over a larger axial range), longer working distances, and often fewer aberrations. Low-NA lenses are thus highly desirable for DT, if we hope to achieve multi-gigavoxel 3D reconstructions in the future. There is thus a tradeoff between the NA of the objective and two desirable properties: multi-gigavoxel SBPs and reduced missing cone artifacts. While rotating the sample~\cite{M_ller_2015}, imaging from multiple angles \cite{konda2018parallelized}, or at multiple focal planes \cite{he2018design} can potentially help fill in missing spatial frequencies, these experimental modifications significantly complicate a standard microscope setup. Thus, unsurprisingly, there is extensive prior work on computational means to fill in the missing cone, such as using positivity constraints \cite{choi2007tomographic, Lim_2015, sung2012stain, sung2011deterministic} and imposing penalties on the spatial image gradient, such as total variation (TV) regularization \cite{Lim_2015, sung2011deterministic,  sung2012stain, krauze2016generalized, delaney1998globally, goris2012electron}. More recently, data-driven deep-learning-based approaches have also been proposed to fill in the analogous missing wedge in X-ray computed tomography (CT) \cite{anirudh2018lose, zhang2016image, ding2019joint}. It has also been shown that accounting for multiple scattering can mitigate the effects of the missing cone (however, this depends on the sample itself exhibiting multiple scattering) \cite{kamilov2015learning, lim2018learning}. Regardless of the forward model employed (e.g., ray-based \cite{choi2007tomographic}, Born\cite{Horstmeyer_2016}, Rytov \cite{Sung_2009}, MS \cite{tian20153d, kamilov2015learning, kamilov2016optical}), or whether or not phase is detected, or the imaging modality itself (e.g., CT, magnetic resonance imaging (MRI), electron microscopy (EM), or standard fluorescence imaging), the missing cone problem is a ubiquitous one and thus addressing it would have far-reaching implications.

In this work, we propose a new approach to address the missing cone problem, termed Deep Prior Diffraction Tomography (DP-DT), which uses a deep image prior (DIP \cite{ulyanov2018deep}) as an \emph{ untrained} deep 3D convolutional neural network (CNN) to generate 3D object reconstructions (Fig. \ref{teaser}). Unlike other recent works that propose to use supervised deep learning to aid in computational image reconstruction problems \cite{dave2018solving, lucas2018using, jo2018quantitative, mccann2017convolutional, barbastathis2019use}, including a number of works that rely on multi-angle illumination \cite{goy2018high, nguyen2018computational, jin2017deep, anirudh2018lose, zhang2016image, ding2019joint, kellman2019data}, the technique proposed here does {\it not} require any pre-training or dataset-specific assumptions. Instead, during iterative object reconstruction, DP-DT simply performs its optimization updates with respect to the parameters of a CNN, as opposed to directly updating the object voxels. The authors of the original DIP paper \cite{ulyanov2018deep} found that the structure of CNNs alone has an inherent bias towards natural images. We thus hypothesized that the artifacts induced by the missing cone problem in the spatial domain would be outside of the domain of natural images that are representable by a DIP. 

Here, we empirically confirm this hypothesis by applying DP-DT to reconstruct 3D images of beads and a biological sample, from their associated variably-illuminated intensity-only images, with higher fidelity than alternative regularization techniques like TV regularization and positivity constraints. We test DP-DT under several different conditions and light propagation models, including the first Born and MS models, and find that the DIP consistently improves 3D reconstruction quality. We also suggest that DP-DT is more general than alternative regularization strategies, as the assumptions of TV regularization and positivity are not always valid, and when they are, they can easily be added into the DP-DT framework. Furthermore, DP-DT does not rely on preexisting datasets, which may be difficult to acquire to learn representations or extract features for object reconstruction. DP-DT thus does not inherit any generalization errors or biases when it is applied to a new types of sample, instead offering a general strategy to improve 3D reconstruction quality.

\begin{figure}[t]
\begin{center}
\includegraphics[width=0.8\columnwidth]{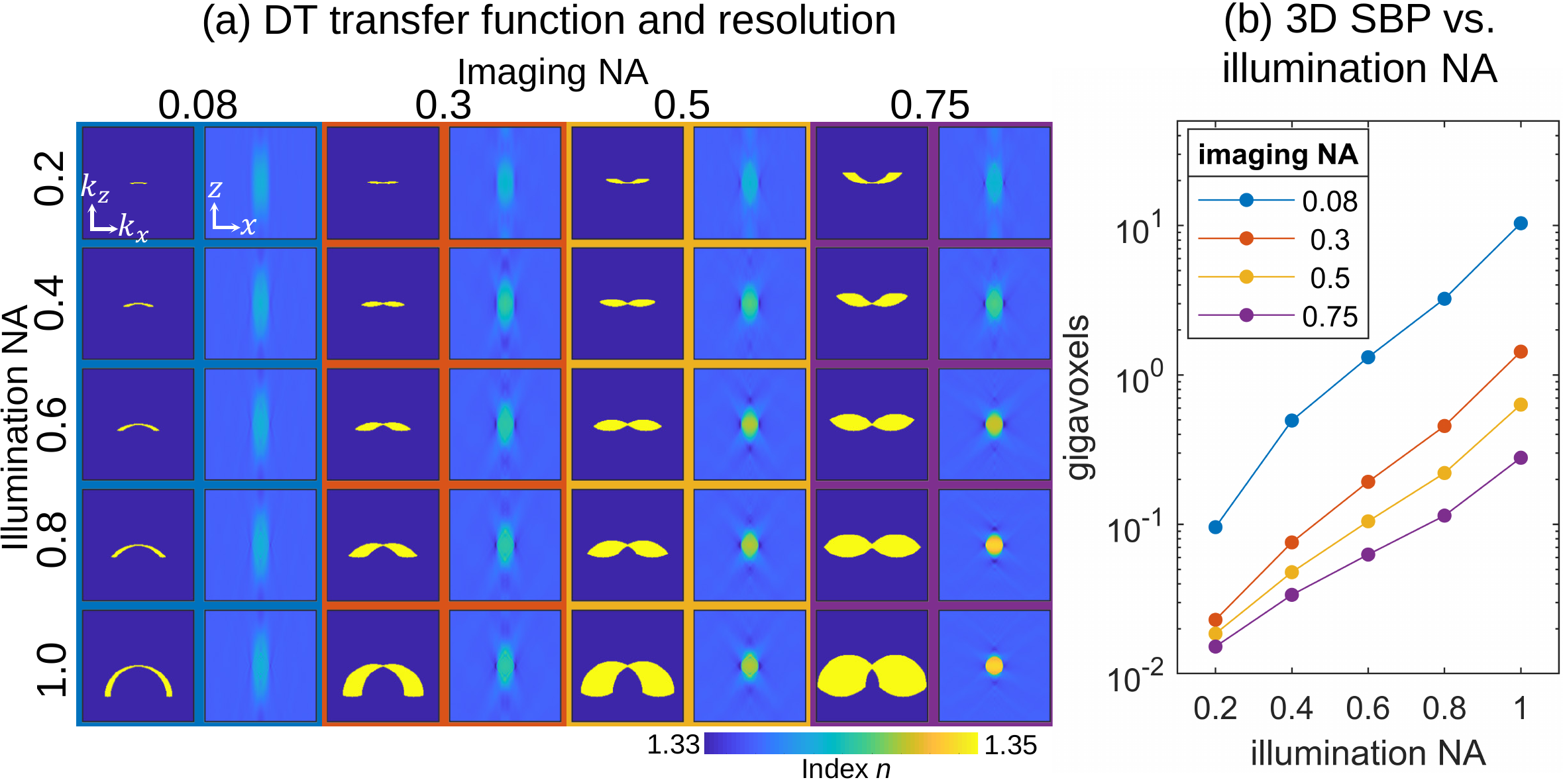}
\caption{The missing cone problem and its effect on the 3D SBP. (a) Simulations of the effects of transfer functions containing missing cones for a variety of illumination (vertical) and collection (horizontal) NAs. The odd columns are the $k$-space supports of the transfer functions and the even columns are the spatial domain representations of 0.8-$\mu m$-diameter bead ($n$=1.35) immersed in water ($n$=1.33), which have been filtered by these transfer functions. (b) 3D SBPs in gigavoxels; 2D fields of view, collection NAs, and magnifications taken from~\cite{Zheng14}; arbitrarily assumes a 20-$\mu m$ axial range; please note the semilog scale.
{\label{missing_cone}}%
}
\end{center}
\end{figure}

\section{Missing cone problem}
To model 3D image formation, we will begin with the first Born approximation, which offers a clear description of the missing cone problem. In general, we can represent the 3D scattering potential of a sample of interest with
\begin{equation}
V(\mathbf{r})=\frac{k^2}{4\pi}\left(n(\textbf{r})^2-n_0^2\right),
\label{V}
\end{equation}
where $\textbf{r}=(x,y,z)$ represents the 3D spatial coordinates of the sample, $k=2\pi/\lambda$ is the vacuum wavenumber, $n_0$ is the surrounding medium's RI, $\lambda$ is the wavelength of light, and $n(\textbf{r})$ is the sample's RI distribution, which is the unknown quantity of interest that we aim to reconstruct. We note that $n(\textbf{r})$ is complex-valued, where $\mathrm{Re}\{n(\textbf{r})\}$ is associated with the index of refraction and $\mathrm{Im}\{n(\textbf{r})\}$ is associated with absorption.

If we take the 3D Fourier transform of $V(\textbf{r})$, we arrive at the sample's scattering potential spectrum:
\begin{equation}
\widetilde{V}(\textbf{k}) = \mathcal{F}_{3D}\left(V(\textbf{r})\right),
\label{V_hat}
\end{equation}
where $\textbf{k}=(k_x, k_y, k_z)$ is the 3D wavevector. Under the first Born approximation, a DT system can only measure a finite range of sample wavevectors, bounded by the angular span of incident and observable light~\cite{Haeberle}. We can represent this limited range of spatial frequencies with a transfer function $H(\textbf{k})$, which defines the observable information at the imaging plane as $\mathcal{F}^{-1}_{3D}(\widetilde{V}(\textbf{k})H(\textbf{k}))$. In practice, $H(\textbf{k})$ can be synthesized by taking a superposition of partial spherical shells (i.e., Ewald spheres), with radii set by the wavelength of light, radial extent set by the imaging NA, and whose positions shift along an arc defined by the illumination k-vector (for mathematical details, see Appendix B). This transfer function's extent in $k$-space shrinks with smaller imaging NAs, even if a large illumination NA is used, resulting in significant blurring in the spatial domain.

To illustrate this point, we modeled the DT transfer function with variable illumination and imaging NAs in Fig. \ref{missing_cone}(a). Next to each DT transfer function, we also simulated its effect on a 0.8-$\mu m$-diameter bead. In these plots, we show 2D $xz$ and $k_xk_z$ slices of the 3D transfer functions. Here, we can see that for a fixed imaging NA, increasing the illumination NA only modestly reduces the axial blurring and RI underestimation induced by the missing cone. Unlike 2D synthetic aperture techniques, such as FP, 3D diffraction tomography methods are thus highly dependent on large imaging NAs for high-fidelity tomographic reconstructions.

As mentioned above, the use of a high imaging NA unfortunately leads to a much smaller lateral imaging FOV. To explore this point in detail, we numerically computed the 3D SBP for each modeled transfer function, which is equivalent to the total number of resolvable voxels in a DT system for a given imaging lens and illumination configuration (Fig. \ref{missing_cone}(b)). In particular, we defined the 3D SBP as the product of the 3D $k$-space volume and the 3D spatial reconstruction volume, basing the SBP computation on the FOV of standard microscope objective lenses~\cite{Zheng14}. Here, we assumed a fixed axial imaging range of 20 $\mu$m. We note that although in theory, even under the first Born approximation, the axial range of the reconstruction volume can be unbounded, in practice the axial range is often limited by factors such as the SNR. In any case, selecting a different axial range only re-scales Fig. \ref{missing_cone}(b)'s y-axis and does not affect the relative comparison between different objective lenses. From this simple analysis, it is clear that lower-NA objectives are desirable for high-throughput tomographic imaging, as they yield significantly larger 3D SBPs. However, such low-NA lenses generate a large missing cone, thus pointing to a critical limitation that must be addressed before it is possible to rapidly acquire high-resolution 3D images.

\begin{figure}[h!]
\begin{center}
\includegraphics[width=.8\columnwidth]{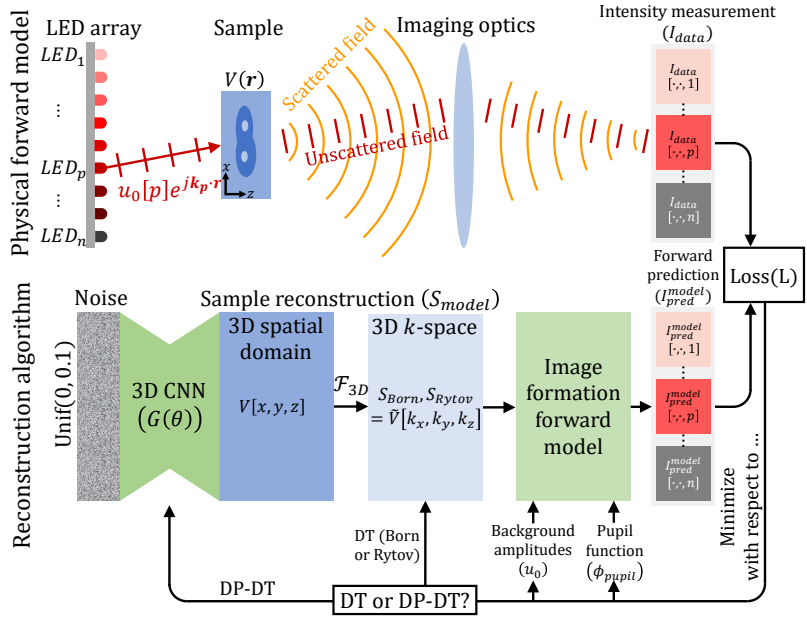}
\caption{{The physical forward model and reconstruction algorithm for DP-DT, under the first Born or Rytov approximation.
{\label{forward_figure}}%
}}
\end{center}
\end{figure}

\section{Deep prior diffraction tomography (DP-DT)}
\label{igw}
To address the above challenge of creating a high-fidelity 3D image reconstruction in the presence of a large missing cone, we propose DP-DT, a forward-model-agnostic framework that merges a deep image prior (DIP) into an iterative tomographic reconstruction process. To demonstrate its versatility, we use DP-DT to improve the quality of 3D sample reconstruction under the assumptions of both the first Born \cite{Horstmeyer_2016} and MS approximations \cite{tian20153d, kamilov2015learning, kamilov2016optical}. While the rest of this section assumes that we are reconstructing with phaseless measurements, we note that the DIP pipeline can easily be applied with different assumed forward models (e.g., \cite{liu2017seagle, lim2019high, kamilov2016recursive}), and can be also applied to phase-sensitive measurements from traditional DT setups, which we include examples of in Appendix C.

\subsection{Inverse problem formulation}
Fig. \ref{forward_figure} shows a high-level summary of the forward image formation process and the inverse problem formulation, with mathematical details presented in Appendix B. In particular, let $S_{model}[\cdot, \cdot, \cdot]$ be the 3D reconstruction target for a particular $model$. For the first Born and Rytov approximations, $S_{Born}$ and $S_{Rytov}$ are a discretized, complex-valued 3D scattering potential spectrum tensor, $\widetilde{V}[\cdot, \cdot, \cdot]$ (Eq. \ref{V_hat}), while for the MS approximation, $S_{MS}$ is a complex-valued 3D tensor, $\delta_{obj}[\cdot, \cdot, \cdot]$ that describes the 3D RI distribution of the sample relative to the medium RI. Next, let the forward predictions for the multi-angle 2D images based on these reconstruction targets be given by $I_{pred}^{model}[\cdot, \cdot, p]$, where $p$ indexes the $p^{th}$ image under illumination from the $p^{th}$ LED and $model$ specifies the employed forward model. Finally, let $I_{data}[\cdot, \cdot, p]$ be a 3D tensor of experimental intensity measurements (i.e., an LED image stack).

The error metric we seek to minimize is the mean square error loss,
\begin{equation}
\label{mse}
E(S_{model}, \phi_{pupil},u_0)=\frac{1}{M}\sum_{i,j,p}\left(\sqrt{I_{pred}^{model}[i,j,p]}-\sqrt{I_{data}[i,j,p]}\right)^2,
\end{equation}
where $\phi_{pupil}$ and $u_0$ are the optionally optimizable pupil phase and per-LED input field amplitudes (see Appendix B), and $M$ is the total number of summed elements. We use the amplitudes rather than the intensities in the loss because it was previously found that the former is more robust to noise and experimental deviations \cite{Yeh_2015}. We may also regularize the reconstruction target to form our final loss function,
\begin{equation}
L(S_{model}, \phi_{pupil},u_0)= E(S_{model}, \phi_{pupil},u_0) + \mathrm{Reg}(S_{model}).
\label{reg_eq}
\end{equation}
We discuss the specific regularization techniques in the next two sections. Thus, the optimization problem for 3D image reconstruction is
\begin{equation}
\argmin_{S_{model},\phi_{pupil},u_0} L(S_{model},\phi_{pupil},u_0).
\label{argmin}
\end{equation}
For the first Born and Rytov approximations, once $S_{model}=\widetilde{V}$ is optimized, we can use Eq. \ref{V} to infer the complex 3D RI distribution.

While we have described intensity-based DT, which is afforded by our LED array setup on a standard microscope, it is straightforward to extend our framework to describe holographic DT setups that also measure phase. The equations to achieve this are presented in Appendix C.

\subsection{Deep image prior (DIP)}
Directly minimizing $L$ can be problematic, given the effects of the missing cone. The DIP is a recently-presented, data-independent method to assist with a large variety of inverse image reconstruction problems without supervised training \cite{ulyanov2018deep, Liu_2019, heckel2018deep, Gong_2018, van2018compressed, mataev2019deepred}. It is an untrained regularization technique that reparameterizes the reconstruction target in the spatial domain as the output of a deep generative CNN that uses pseudorandom noise as input (see Fig. \ref{DIP} in Appendix A for the architecture we used). After initializing the CNN with pseudorandom noise, DIP optimization then proceeds to update CNN weights to minimize loss, as opposed to directly optimizing the reconstruction target ($S_{model}$). Here, we hypothesize that the DIP's resistance to unnatural images extends to the third dimension and can help eliminate missing cone artifacts in diffraction tomography.

Compared to the DIP, other recently proposed techniques based on deep neural networks for FP reconstruction \cite{shamshad2019deep, Nguyen_2018, Kappeler_2017, icsil2019deep, metzler2018prdeep} all rely on pretraining, and hence are inherently biased towards a particular set of examples found in the training data set. Furthermore, these techniques were primarily applied to 2D reconstruction. However, while the 2D FP inverse problem is well-posed if the LEDs are sufficiently densely packed (i.e., to obtain $>50\%$ overlap \cite{bunk2008influence}), the 3D FP inverse problem is always ill-posed due to the missing cone, no matter how densely packed the LEDs are. Hence, the 3D inverse problem presents a more significant challenge that could benefit more significantly from the DIP.

To incorporate the DIP reparameterization into our framework, we modify Eq. \ref{argmin}, for the first Born (or Rytov) and MS forward models, respectively:

\begin{equation}
 \argmin_{\theta,\phi_{pupil},u_0} L_{Born}(\widetilde{V} = \mathcal{F}_{3D}\left(G(\theta)\right),\phi_{pupil},u_0),
\label{DIP_final_born}
\end{equation}
\begin{equation}
 \argmin_{\theta,\phi_{pupil},u_0} L_{MS}(\delta_{obj} = G(\theta),\phi_{pupil},u_0),
\label{DIP_final_MS}
\end{equation}
where $G$ is deep generative CNN parameterized by $\theta$. Eqs. \ref{DIP_final_born} and \ref{DIP_final_MS} are the final optimization problems for DP-DT. Note that $G$ outputs a prediction for the 3D object in the spatial domain, and thus we take its Fourier transform to get $\widetilde{V}$ in Eq. \ref{DIP_final_born}. However, regardless of the form of $S_{model}$, it is critical to note that the optimization now is performed with respect to $\theta$, which are the weights within the CNN.

\subsection{Other regularization}
It is straightforward to also include other well-known regularization methods in Eq.~\ref{reg_eq}, in particular TV and positivity regularization, to which we compare our proposed approach. We used the isotropic TV regularization, given by
\begin{equation}
R_{TV}(S(\mathbf{r}))=\sum_{\mathbf{r}}\sqrt{|\nabla_xS(\mathbf{r})|^2+|\nabla_yS(\mathbf{r})|^2+|\nabla_zS(\mathbf{r})|^2},
\end{equation}
where spatial gradients are approximated by finite differences. TV regularization promotes piecewise smoothness by encouraging sparsity in reconstruction gradients.

The positivity constraint that we test is applied to the real part of the RI under the assumption that the sample index does not fall below that of the immersion medium, noting in Fig. \ref{missing_cone} that one manifestation of the missing cone problem is negative index value artifacts:
\begin{equation}
R_+(n(\mathbf{r}))=\sum_{\mathbf{r}}\text{min}(\text{Re}\{n(\mathbf{r})\}-n_0, 0)^2.
\end{equation}
This expression returns 0 when the real part of the RI is above the immersion RI, and a quadratic penalty otherwise.

Note that both of these regularization terms are differentiable almost everywhere and thus are suitable for gradient-based optimizers. 
The modified loss function to be minimized is thus
\begin{equation}
L=E+\lambda_{TV}R_{TV}+\lambda_+R_+,
\end{equation}
where $\lambda_{TV}$ and $\lambda_+$ are regularization coefficients tuning the respective relative contributions. Unless otherwise specified, the DIP-based reconstructions do not include these extra positivity or TV regularization terms.

\section{Results}

\subsection{Setup for experiments and simulations}
In this section, we demonstrate the effectiveness of DP-DT in both simulation and experiment. Our experimental setup (previously described in Ref. \cite{Horstmeyer_2016}) consists of a standard microscope equipped with an infinity-corrected 20$\times$ objective lens (NA=0.4, Olympus MPLN), an 8-bit camera with 1920$\times$1456 4.54-$\mu m$ pixels (Prosilica GX 1920), and a 31$\times$31 LED array as the illumination source (SMD3528, center wavelength = 632 nm, 4-mm LED pitch), positioned below the sample to give an illumination NA of approximately 0.4. In our simulations, we used the same setup, but varied the distance between the LED board and the sample to tune the illumination NA, and varied the diameter of the aperture function ($A[\cdot,\cdot]$, see Appendix B) to tune the imaging NA. For experimental results, we ignored the dark-field images from the corners of the LED array due to a low SNR, resulting in a total of 641 multi-LED images. However, for simulations, we used all 31$\times$31 LEDs in a centered square grid, computationally increasing the exposure (or equivalently, the illumination intensity) for the dark-field LEDs to boost the detected image SNR. To make simulations more realistic, we added Poisson-distributed noise to the forward intensity predictions, assuming a pixel well depth of 50,000 photoelectrons, and discretized the result into 8 bits. To ensure that the regularizers are only accounting for the missing cone, we simulated samples that approximately followed the first Born approximation and used this for both simulation and reconstruction. We also note that jointly optimizing the pupil function did not make much of a difference in terms of the reconstruction quality, and we thus did not end up doing so for either the simulated or experimental results.

We specified the forward model and performed gradient descent using the Adam optimizer \cite{kingma2014adam} in TensorFlow \cite{abadi2016tensorflow}. The code will be available at deepimaging.io. To accelerate optimization, we used a NVIDIA Tesla T4 GPU on the Google Cloud Platform. By default, we use the entire dataset per optimization step if it fit within the GPU's 16 GB of memory. Otherwise, we split the dataset into roughly evenly-sized batches along the LED dimension such that the batches each fit in memory. After every full pass through the dataset (i.e., an epoch), we reshuffled the data along the LED dimension before splitting into a new set of batches. Optimization times depend on the FOV, the batch size, whether DIP is used (DIP requires several times more iterations), and which scattering model is used (the MS model is more computationally expensive than the first Born model). For our experimental implementation, each iteration was on the order of seconds, resulting in optimization times on the order of minutes to an hour (100s to 1000s of iterations) without DIP, and several hours (10s of 1000s of iterations) with DIP.

\begin{figure}[!t]
\begin{center}
\includegraphics{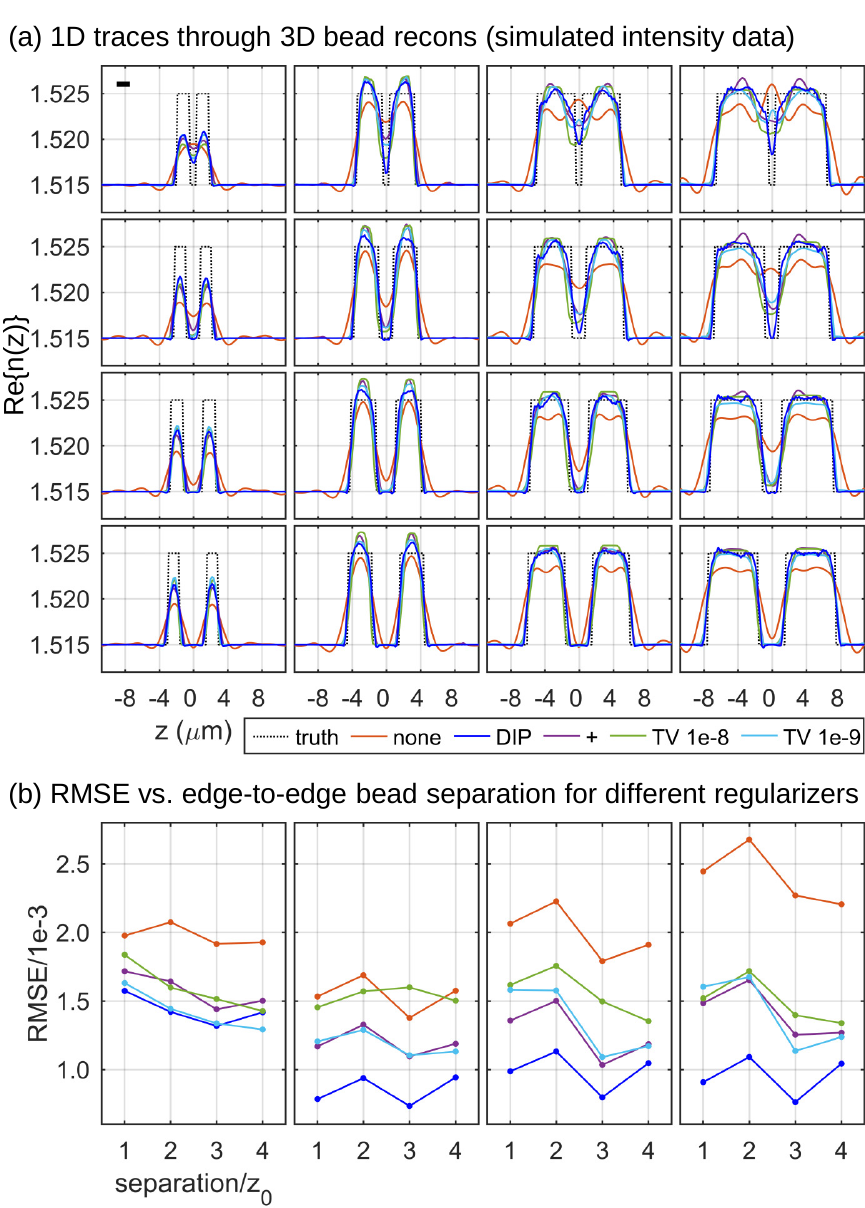}
\caption{Comparison of 3D reconstruction quality for phaseless DT using several regularizers. (a) 1D traces through simulations of two beads spaced axially. Rows show different bead axial separations, while columns show different bead sizes (imaging NA = 0.4, illumination NA = 0.5). Each curve corresponds to a different regularization technique (i.e., none, DIP, positivity (+), and TV) and ground truth. Scale bar corresponds to the Nyquist period. (b) The RI RMSEs from the ground truths for each regularizer. Each of the four plots corresponds to a different bead size, and each curve corresponds to a different edge-to-edge bead separation, where $z_0=$ 0.75 $\mu m$.}
\label{bead_sim}%
\end{center}
\end{figure}

\begin{figure}[!ht]
\begin{center}
\includegraphics{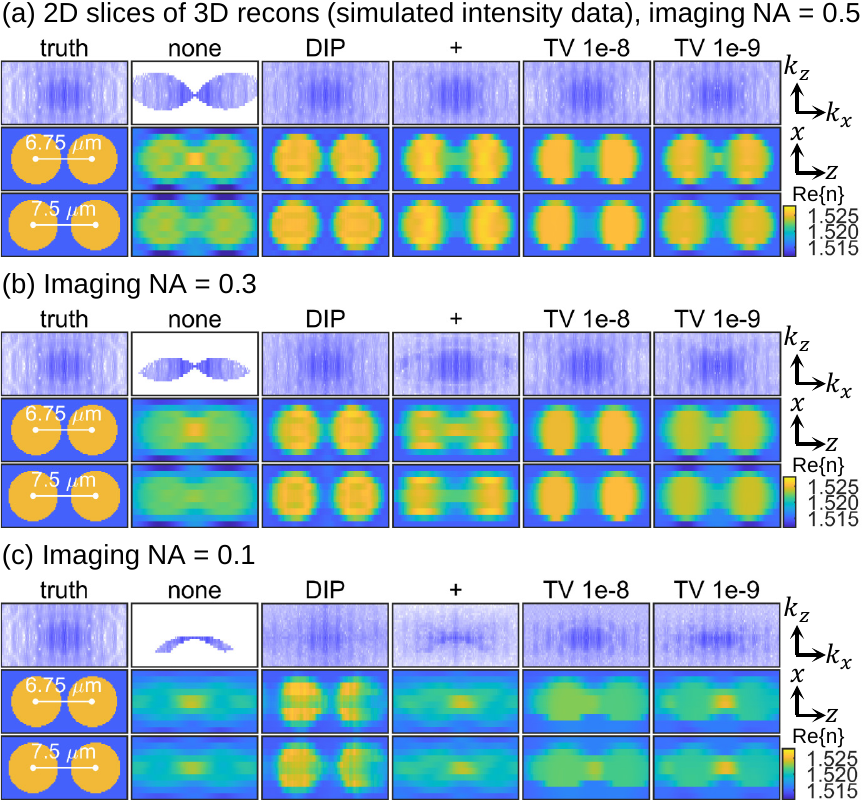}
\caption{2D cross-sections of select simulated bead pairs. (a) The first row is the through-origin $k_xk_y$ cross-sections of the scattering potential spectra of the reconstructions containing all the bead pairs, under illumination NA = 0.4, imaging NA = 0.5. The second and third rows show through-center $xz$ cross sections at two different separations. (b) The same information as (a), but under imaging NA = 0.3. (c) The same information as (a), but under imaging NA = 0.1.}
\label{bead_sim2D}
\end{center}
\end{figure}

\subsection{Bead simulation results}
First, to test the axial resolution of DP-DT, we simulated 31$\times$31 intensity-only images under the first Born approximation from pairs of 3D beads ($n$=1.525, on a background of $n_0$=1.515) of various sizes and various axial spacings under illumination from a spatially coherent LED array. We tested multiple imaging NAs (0.1, 0.3, 0.5) and a fixed illumination NA of 0.4. We reconstructed the 3D RI map of the bead phantom using the following priors: none, DIP, positivity, TV ($\lambda_{TV}$=1e-8), and TV ($\lambda_{TV}$=1e-9). Fig. \ref{bead_sim} summarizes these results for a 0.5 imaging NA, showing 1D axial RI cross-sections through the center of the 3D bead reconstructions (Fig. \ref{bead_sim}(a)), as well as the root mean square error (RMSE) of these 1D axial profiles with respect to the ground truth (Fig. \ref{bead_sim}(b)). We can see that DP-DT performs as well as or better than the other regularizers. In particular, in all four columns of Fig.\ref{bead_sim}(a), there are separations for which the dip in between the two beads is deeper for DP-DT than the other regularization strategies, and in almost all cases DP-DT has a smaller RMSE than the other regularized results (Fig. \ref{bead_sim}(b)). Similar observations apply to the imaging NAs of 0.1 and 0.3 (Supplementary Figs. \ref{bead_sim_0_1} and \ref{bead_sim_0_3} in Appendix C). Furthermore, we obtained very similar results for imaging NAs of 0.1, 0.3, and 0.5 in the case of reconstructing from phase-sensitive measurements (Supplementary Figs. \ref{bead_sim_phase_0_1}, \ref{bead_sim_phase_0_3}, and \ref{bead_sim_phase_0_5} in Appendix C).

Fig. \ref{bead_sim2D} shows select 2D $xz$ RI cross-sections for the largest particular bead size and the two closest separations, from which we can see that the gaps between the two beads for DP-DT is more pronounced and faithful to the ground truth, compared to the other regularization techniques. Note that even though this sample is ideally suited for a TV prior, as it is piecewise smooth, DP-DT still produces superior results. We obtained similar results for reconstructions from phase-sensitive measurements (see details in Appendix C). We emphasize that for DP-DT the network was not pretrained to prefer such reconstructions; rather, these results were innately preferred by the DIP's CNN structure. 

\begin{figure}[t!]
\begin{center}
\includegraphics[width=.8\columnwidth]{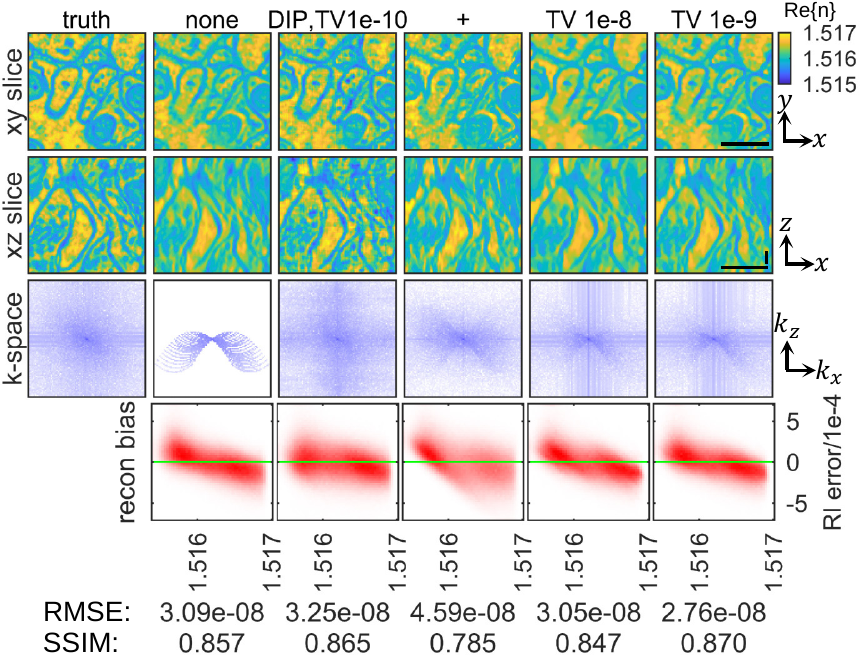}
\caption{Simulated biological sample, reconstructed under various regularizers. The first row shows an $xy$ cross-section of the reconstruction volume, the second row shows and $xz$ cross-section, and the third row shows the through-origin $k_xk_y$ cross-section of the scattering potential spectrum. The fourth row shows 2D histograms, where the vertical axis is the RI error from the ground truth and the horizontal axis is the ground truth RI. Scale bars, 15 $\mu m$.}.
\label{EM}
\end{center}
\end{figure}


\subsection{Biological simulation results}
It is expected that TV-regularized reconstructions would perform well for bead samples, because such samples contain regions of uniform RIs. Thus, we also simulated a more complicated and realistic biological sample, based upon 3D isotropic EM images of hippocampal cells \cite{Lucchi_2012}, for which the smoothness imposed by TV regularization may not be as appropriate an assumption. To convert the 3D EM data into a ground-truth 3D RI map, we renormalized its voxelized measurements to extend from 1.515 (i.e., $n_0$) to 1.517. With this rescaled dataset as the 3D RI, we used the first Born approximation to simulate 2D multi-LED intensity images, using an imaging NA of 0.2 and illumination NA of 0.4. The reconstruction results, assuming the same priors as used for the bead simulation (none, DIP+TV 1e-10, positivity, TV 1e-8, and TV 1e-9) are shown in Fig. \ref{EM}. Note that the DIP reconstruction contains a small amount of TV regularization, which we found produced better results than DIP alone \cite{Liu_2019}. While in the $xy$ dimensions (Fig. \ref{EM}, first row), the reconstructions look similar across the different priors, in the $xz$ dimensions (Fig. \ref{EM}, second row), TV regularization only axially blurs the reconstruction. Furthermore, positivity regularization does not offer much improvement over the non-regularized reconstruction. However, DP-DT is able to estimate a significant amount of information in the missing cone, which is also apparent when considering the $k_xk_z$ cross-sections of the scattering potential spectra (Fig. \ref{EM}, third row). 
\begin{figure}[t!]
\begin{center}
\includegraphics[width=.8\columnwidth]{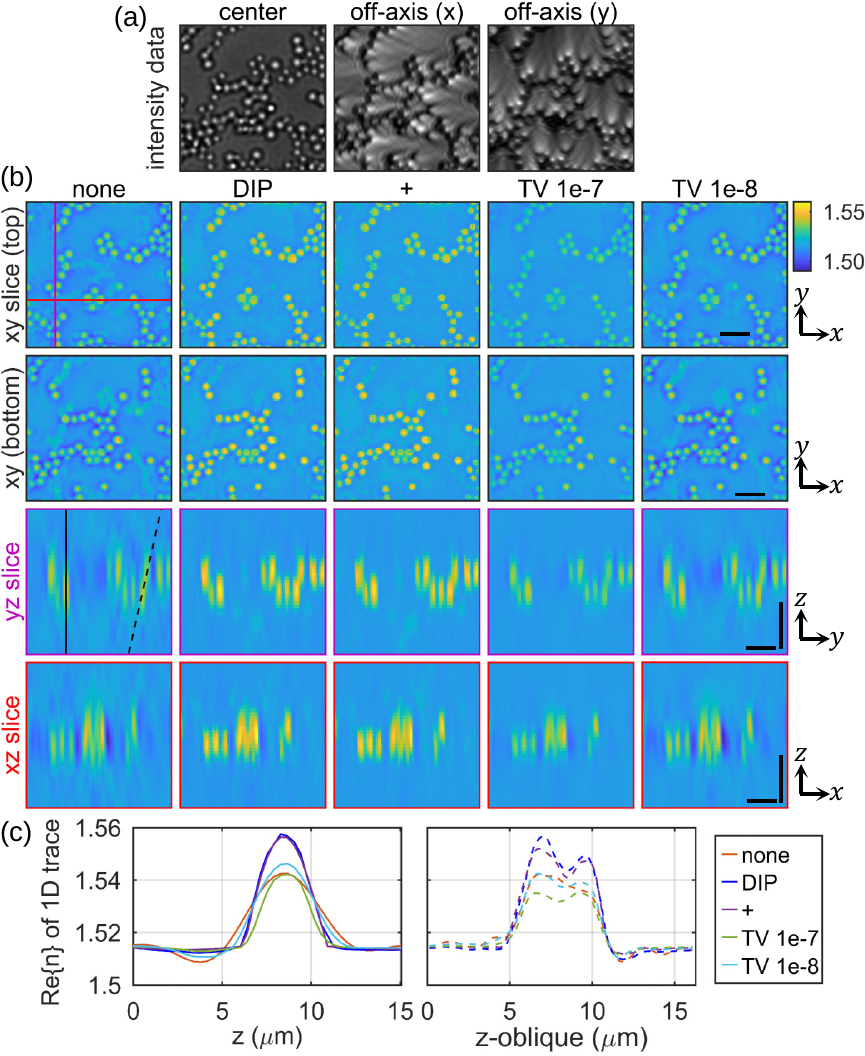}
\caption{Experimental 2-layer, 2-$\mu$m bead results, using the first-Born model. (a) Example raw LED images. (b) Comparison of performance of different regularizers (columns). The first two rows are $xy$ slices through the two bead layers, and the last two rows are two axial slices whose positions are indicated by the red and magenta lines in the upper left plot. Scale bars, 5 $\mu m$. (c) 1D traces through regions indicated in the first panel of the $yz$ slice row. The expected RI of the beads is 1.59.}
\label{2layer}
\end{center}
\end{figure}

To quantify the comparisons of the different techniques, we computed the RMSE and the structural similarity (SSIM) index \cite{wang2004image}, which are displayed at the bottom of Fig. \ref{EM}. While DP-DT did not produce the best statistics, its reconstruction visually exhibits less blurring and fewer artifacts. We also found that DP-DT produces less biased estimates of RI compared to all the other techniques (Fig. \ref{EM}, last row). In particular, while the non-DIP approaches exhibit reconstructions that overestimate RI for low RI values, and underestimate RI for high RI values (as one would expect for spatial blurring), DP-DT produces unbiased estimates for low RI values and less biased results for high RI values (i.e., its flatter red histogram indicates a more consistent bias-variance tradeoff with respect to RI). This is a unique property of DIP, as most conventional regularization techniques trade off unbiasedness for lower variance in their estimations.

\subsection{Experimental results, first Born approximation}
To experimentally test DP-DT with phaseless 3D imaging data, we first examined its imaging performance using two bead phantom samples. We first created a bead phantom consisting of two layers of 2-$\mu$m-diameter beads ($n$=1.59), separated by 3.9 $\mu$m and embedded in oil ($n_0$=1.515). We reconstructed the 3D RI map of this two-layer sample under the first Born approximation, using no prior, DIP, positivity, TV ($\lambda$=1e-7), and TV ($\lambda$=1e-8). The top two rows of Fig. \ref{2layer}(b) shows two $xy$ cross-sections at the two layers of interest. Without regularization, there is leakage of information between layers, preventing clean separation, due to the missing cone artifact. This is also apparent in the $yz$ and $xz$ slices (Fig. \ref{2layer}(b), bottom two rows). While all of the regularized reconstructions were effective to some extent in reducing this artifact, reconstructions with no regularization and with TV show severely underestimated bead RIs. DIP- and positivity-regularized reconstructions were less underestimated. RI underestimation is seen more clearly in Fig. \ref{2layer}(c), which show 1D traces in the $yz$ plane, among with DP-DT shows the least underestimation. The remaining RI underestimation may be explained by the small bead sizes, noting that there is also RI underestimation in the reconstructions of our simulated beads in the first column of Fig. \ref{bead_sim}, even for DP-DT (also note that for the closest separations, DP-DT has the lowest degree of RI underestimation). Also, the beads in Fig. \ref{2layer} are not axially resolved fully, consistent with the incomplete separation in the upper left panel of Fig. \ref{bead_sim}. Differences between our simulated and experimental results (i.e., better axial separation in simulation) may be attributed to  imperfections in our experimental setup, most notably the challenge of establishing an exact correspondence between the estimated and true LED positions, and perhaps the assumed scattering model. As shown in section \ref{results_MS} below, DP-DT using the MS model both reduces the degree of RI underestimation and improves the separation between the two bead layers.

\begin{figure}[t!]
\begin{center}
\includegraphics[width=.8\columnwidth]{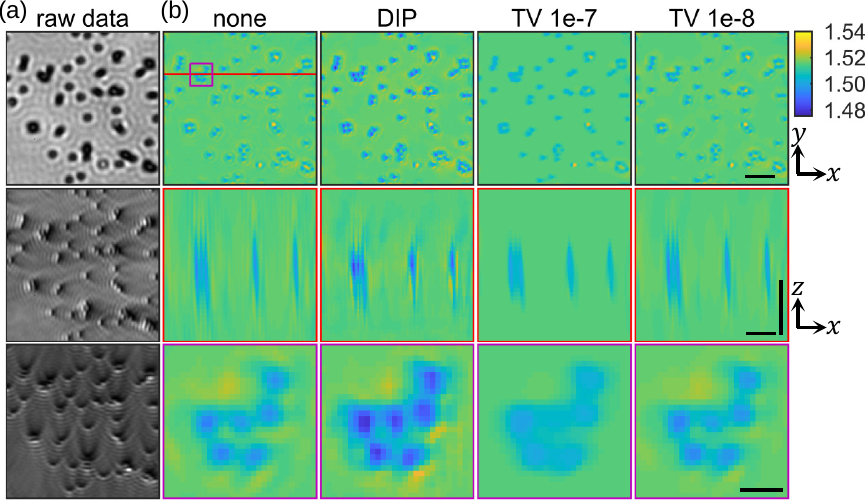}
\caption{Experimental 800-nm bead sample. (a) Example raw LED images. (b) Comparison of different regularizers, excluding positivity because the RI is below that of the medium. The first row is $xy$ cross-sections and the second row is $xz$ cross-sections indicated by the red line. The third row shows a close-up view of the first row, indicated by the purple box. Scale bars, 5 $\mu m$ (top two rows of (b)), 1 $\mu m$ (bottom row of (b)).}
\label{small_bead}
\end{center}
\end{figure}

\begin{figure}[t]
\begin{center}
\includegraphics[width=.8\columnwidth]{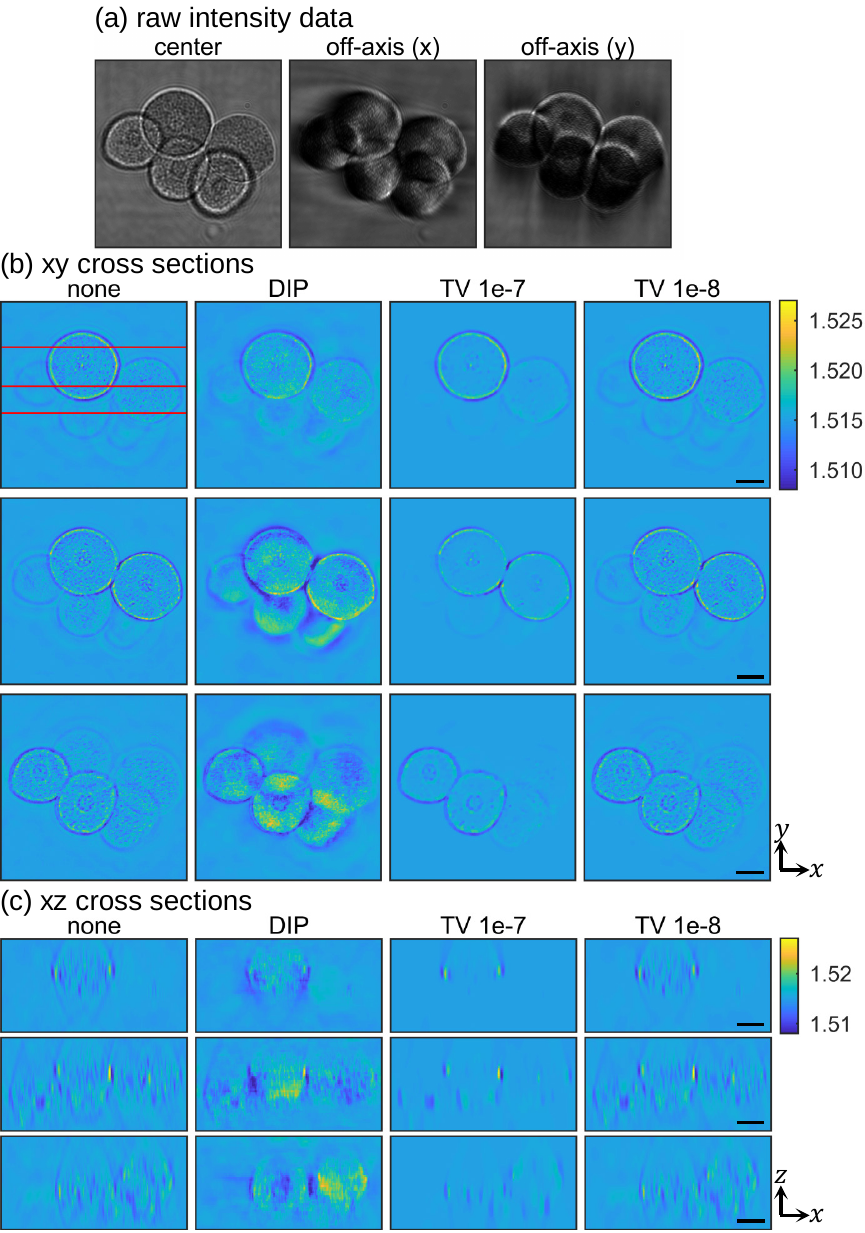}
\caption{Experimental stacked starfish embryo results. (a) Example raw LED images. (b) $xy$ cross-sections at various depths. (c) $xz$ cross-sections at positions indicated by the horizontal red lines in the upper left panel of (b). Scale bars, 10 $\mu m$.}
\label{starfish}
\end{center}
\end{figure}

Our second experimental target consisted of a single layer of 800-nm-diameter beads with an unspecified RI below that of the embedding oil ($n_0$=1.515). We reconstructed the 3D RI map using the same priors as for the 2-layer bead sample (Fig. \ref{small_bead}), except for positivity, because the 800-nm bead samples exhibited RI values lower than that of the medium. Note the enhanced lateral resolution in all reconstructions with respect to the raw data, as expected via DT aperture synthesis. The more heavily TV-regularized  result ($\lambda_{TV}$=1e-7), while exhibiting reduced axial missing cone artifacts, has a reduced lateral resolution compared to the other reconstructions (Fig. \ref{small_bead}(b), bottom row). Furthermore, as with the 2-layer sample, we found that the RI difference estimate for the unregularized and the TV-regularized reconstructions were underestimated, unlike for the DP-DT result (Fig. \ref{small_bead}). Finally, we note that the DP-DT reconstruction contains more energy concentrated at the center of the beads along the axial dimension, demonstrating a successful counter to the effects of the missing cone.

As the final sample, we imaged fixed early-stage starfish embryo cells with an LED-outfitted microscope (same configuration as for the bead experiments, Fig. \ref{starfish}). We used the same regularization settings as for the 800-nm bead sample (also without positivity regularization, as we were not certain that the RI was strictly greater than that of the medium). $xy$ and $xz$ slices of the reconstructions are shown in Fig. \ref{starfish}, respectively. Here, we can see that TV removes axial artifacts due to the missing cone problem, but at the cost of blurring features in the lateral and axial dimensions and even erasing many of the cellular features within each embryo. This is because the piecewise smoothness assumption of TV may not be appropriate for this biological sample with a relatively highly varying spatial RI profile. On the other hand, the DP-DT reconstruction not only has a higher axial resolution, but also produces cells with a rounder appearance in the axial direction, while the other reconstructions exhibit characteristic missing cone artifacts that cause the cells to taper in the axial direction. It is worth noting that DP-DT produces higher RI estimates, which is further evidence that DP-DT is filling in the missing cone, which would otherwise cause RI underestimation.

\begin{figure}[t!]
\begin{center}
\includegraphics[width=.8\columnwidth]{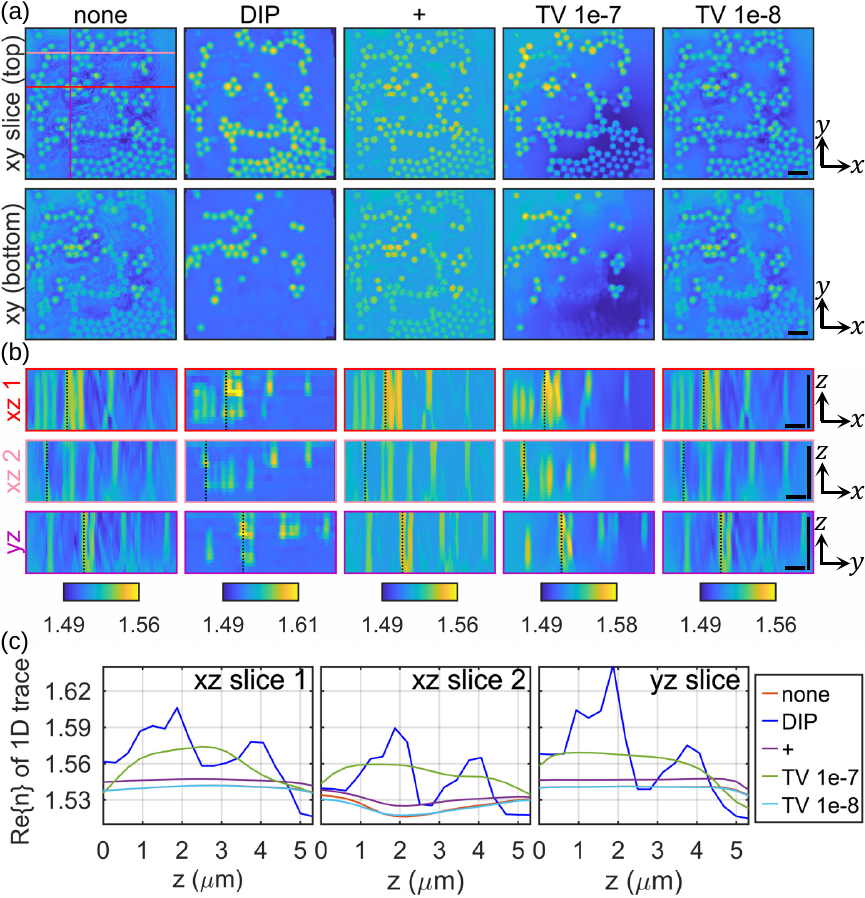}
\caption{Experimental 2-layer, 2-$\mu$m bead results, using the multi-slice model. (a) $xy$ slices at the two bead layers. (b) Axial slices indicated by the lines in the upper left panel of (a). (c) 1D axial traces through the dotted lines in (b) (expected RI=1.59). Scale bars, 5 $\mu m$.}
\label{2layer_MS}
\end{center}
\end{figure}

\subsection{Experimental results, multi-slice (MS) approximation}
\label{results_MS}
We also tested DP-DT under the MS forward model with the experimental 2-layer, 2-$\mu$m bead phantom. Because the MS forward model is more computationally intensive, we used what we call ``spatial patching'', whereby at each iteration we select a random, apodized spatial crop within the reconstruction over which to optimize (for more details, see Appendix B). We also note that for the cases of positivity, weaker TV (1e-8), and no regularization, we had to terminate the optimization early to prevent the reconstruction algorithm from diverging.

The results are shown in Fig. \ref{2layer_MS}. Without regularization, or with positivity or weak TV regularization, the axial resolution is very poor, with poor discrimination of the two layers. However, DP-DT here has superior axial resolution, even resolving the beads from the two layers, which was not possible with DP-DT with the first Born model. Furthermore, the RI values reconstructed using DP-DT exhibit the least underestimation, approaching the expected value of 1.59. These results may be attributed to the fact that the MS model is able to model multiple forward scattering events; however, note that only DP-DT attains this RI value and axial separation.

\section{Discussion and conclusion}

In summary, we have presented DP-DT, a flexible and general framework that augments existing 3D diffraction tomography techniques with a DIP, which we have shown to alleviate the effects of the missing cone problem. Specifically, we have applied DP-DT to two scattering models, the first Born and MS approximations, and demonstrated its effectiveness in simulated and experimental data with intensity measurements, as well as with simulated phase-sensitive measurements. DP-DT differs from other deep-learning-based approaches in that it does not require pre-training on and hence is not biased towards a pre-existing dataset. DP-DT can thus be applied in situations where it is expensive or otherwise infeasible to collect large datasets for supervised training. Instead, DP-DT relies on the inherent preference of CNN structures for ``natural'' images, a class of images which we have empirically shown to exclude images with missing cone artifacts. These results open the door to 3D DT with multi-gigavoxel-scale SBPs. 

 We used a single common architecture for all the reconstructions in this paper (see Appendix A), based on the recently reported encoder-decoder DIP architecture \cite{ulyanov2018deep}, which we did not have to tune for specific samples. Note that the number of parameters in this architecture is fixed, regardless of the reconstruction size, because it is a fully convolutional network that adapts to the output size. Unlike with conventional priors, whose impact may be tuned by a coefficient, there is no obvious way to adjust the ``degree of naturalness'' conferred to the reconstruction by the DIP, except perhaps indirectly by the CNN's architecture. Thus, future work may include exploring the impact of alternative architectures \cite{heckel2018deep}.

Furthermore, we did not have to employ early stopping to avoid overfitting for DP-DT, unlike in the original paper \cite{ulyanov2018deep}, which found that running the optimization for too many iterations resulted in recapitulation of image artifacts. We hypothesize that this may be because DP-DT indirectly inpaints in the Fourier domain, such that the missing information is not spatially localized. However, it is possible that the MS model may experience overfitting, as it is not explicitly a Fourier inpainting approach (outside of the weak-scattering limit, in which it coincides with the first Born model). We did not, however, run the DIP-regularized MS optimization loops long enough to observe such effects.

We found that during optimization of DIP-regularized reconstructions under the first-Born approximation, the optimization would sometimes diverge rapidly, effectively resetting the reconstruction. The authors of the original DIP paper also observed a similar phenomenon. To counteract this divergence in an automated fashion, we periodically checkpointed the parameters and monitored the ratio of the current loss versus the mean loss over the last few iterations. If the ratio exceeded a certain threshold, we would restore the parameters to the previous checkpoint and anneal the learning rate by a factor of 0.9. We note that we did not observe this phenomenon for DIP-regularized MS reconstructions, though it is unclear whether that was due to the scattering model or the fact that we used spatial patching.

Future work includes further investigation of DP-DT under the MS forward model, especially when scaling to larger FOVs. Due to memory constraints, we would have to use smaller LED batch sizes and smaller spatial patch sizes relative to the full FOV. The DP-DT reconstruction under the MS model also showed artifactually high RI values near the axial edges of the reconstruction volume. However, these are not of concern because they occurred outside of the object support. These high values are not seen in non-DIP reconstructions, perhaps because they were initialized at 0, while DIP does not default to 0 (otherwise, it wouldn't fill in the missing cone). Other future work includes investigating whether the DIP can account for artifacts that arise from using a scattering model in situations where its assumptions are not met, or artifacts from experimental uncertainties. Finally, although our primary goal is to achieve high fidelity multi-gigavoxel-scale 3D image reconstructions, currently, the 3D FOVs that we reconstructed here are on the order of 10s of $\mu m$ in the lateral dimensions. To achieve such SBPs, we would thus need to reconstruct the full FOV afforded by low-NA objectives, which can be on the order of several mm. In other words, we are currently about two orders of magnitude away in terms of a full demonstration. The major challenge of scaling up to multi-gigavoxel imaging using DP-DT is computation time, whether we use spatial patching or just divide up the reconstruction into patches and reconstruct them sequentially. Thus, we will explore more memory-compact DIP architectures and will investigate the effect of batch size and density of illumination angles in order to make the reconstruction process more computationally tractable.

While we have so far demonstrated DP-DT for coherent diffraction tomography, utilizing two popular scattering models and various samples, we expect our results to be more generally applicable to other imaging modalities that exhibit artifacts due to the missing cone problem. This includes ill-posed problems \cite{van2018compressed} and other domain-specific problems (e.g., anisotropy, speckle, coherent ringing artifacts, noise, etc.) from the fields of 3D X-ray CT, MRI, EM and fluorescence microscopy. Furthermore, DP-DT may certainly be applied to other scattering models, such as higher-order Born approximations and other multiple scattering models that may exhibit unstable convergence during optimization that would invariably produce unnatural-looking reconstructions, against which DP-DT may safeguard. We already observed this optimization stabilizing phenomenon for the MS-model-based reconstructions, where without DIP or heavy TV regularization the reconstruction acquired extreme artifacts, similar to the unregularzied reconstructions in Fig. S3(e),(f) in Chowdhury et al. 2019 \cite{chowdhury2019high}. Finally, we also hope to apply DP-DT to setups that include reflection geometries \cite{roberts1985reflection,zhou2019optical}, which invariably contain gaps in $k$-space between disjoint high and low frequency bands. In conclusion, we are hopeful that DP-DT will open up options for using wider-FOV, lower-NA imaging lenses for 3D imaging without axial reconstruction artifacts, thus paving the way for multi-gigavoxel tomographic imaging in the future.

\section*{Funding} KCZ was supported by the National Science Foundation (DGF-1106401) and RH was supported by funding from the Erlangen Graduate School in Advanced Optical Technologies (SAOT) by the German Research Foundation (DFG) in the framework of the German excellence initiative.

\section*{Acknowledgments} The authors would like to thank the lab of Prof. Changhuei Yang for assistance with experimental data capture, as well as Ruobing Qian, Shiqi Xu, and Mykola Kadobianskyi for helpful comments during preparation of this manuscript.

\section*{Disclosures} The authors declare no conflicts of interest.

\section*{Appendix A: CNN architecture}
\label{architecture}
\begin{figure}[t!]
\begin{center}
\includegraphics[width=.8\columnwidth]{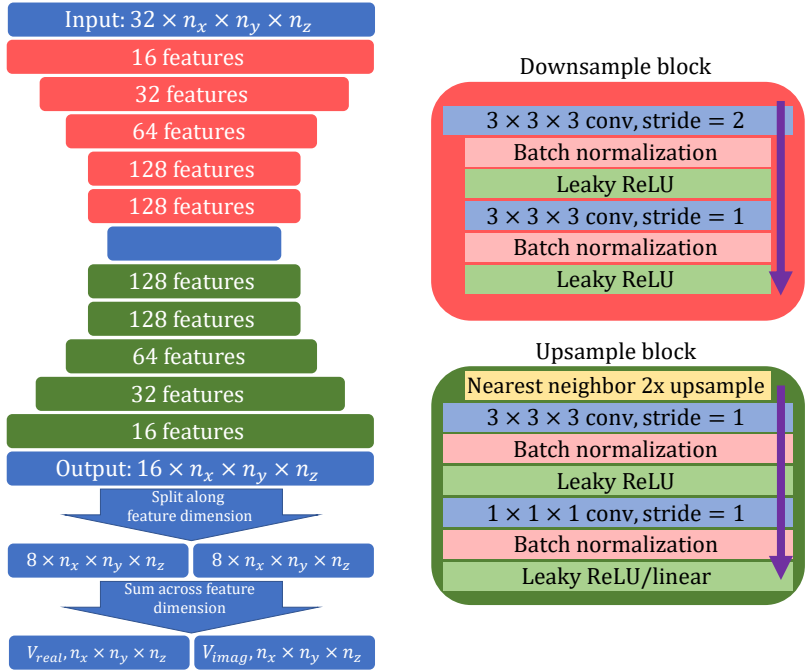}
\caption{{The DIP architecture used for the DP-DT reconstructions in this paper.
{\label{DIP}}
}}
\end{center}
\end{figure}

For our DIP's convolutional neural network structure, we adopted a slightly modified version of the symmetric encoder-decoder architecture used in the original paper \cite{ulyanov2018deep}, detailed in Fig. \ref{DIP}. The input is sampled from a uniform random distribution between 0 and 0.1, and is fixed throughout optimization. The downsampling blocks used strided convolutions as the downsampling operation, while the upsampling blocks used nearest neighbor upsampling. Unlike the original paper, we did not use skip connections. The original network made heavy used of batch normalization \cite{ioffe2015batch} and leaky ReLU activations \cite{xu2015empirical}. To generate the real and imaginary components of $V$, we split the real-valued output of $G$ into two equally sized tensors and summed across the feature dimension. Since leaky ReLU has a preference for positive numbers, we used a linear activation in the final upsampling block if we do not necessarily expect the sample scattering potential to be strictly positive.

\section*{Appendix B: Forward models}
\label{forward_models_detailed}
\subsection*{First Born and Rytov approximations}
\label{born_rytov}
To provide a mathematical description of our forward model that is amenable to computational reconstruction, we assume discretized coordinates where appropriate to emphasize practical implementation. First, we define the following collection of wavevectors,
\begin{equation}
\mathbf{k^{cap}}[i,j]=(k^{cap}_x[i,j], k^{cap}_y[i,j], k^{cap}_z[i,j]),
\end{equation}
where the $[i,j]^{th}$ wavevector is of a partial spherical ``cap,'' which is a segment of a discretized sphere in $k$-space (i.e., an Ewald sphere) with radius $k$. We initially place the center of this spherical cap at the origin of $k$-space at $[i,j]=[0,0]$. The $[i,j]$ indexing emphasizes that the spherical cap, although defined in 3D $k$-space, is indexed on a 2D Cartesian grid when orthographically projected onto the $k_xk_y$ plane. The maximum lateral extent of the spherical cap is defined by the NA of the imaging lens such that $k_{xy}^{max}=kN\!A$, where $N\!A$ is the imaging numerical aperture. Next, let
\begin{equation}
\mathbf{k^{ill}}[p]=(k^{ill}_x[p], k^{ill}_y[p], k^{ill}_z[p])
\end{equation}
be the $p^{th}$ illumination wavevector corresponding to the position of the $p^{th}$ LED relative to the sample. Subtracting the illumination vector from the spherical cap coordinate vector places the spherical cap coordinate in the correct location in 3D $k$-space for the $p^{th}$ LED. Following the Fourier diffraction theorem \cite{Wolf_1969}, the $[i,j]^{th}$ coordinate of the 2D discrete Fourier transform (DFT) of the field scattered off of the object from the $p^{th}$ illumination, measured at the detector plane, is provisionally,
\begin{equation}
\widetilde{u}[i,j,p]=\frac{\widetilde{V}(\mathbf{k^{cap}}[i,j]-\mathbf{k^{ill}}[p])}{4\pi1ik_z^{cap}[i,j]}
\label{FDT}
\end{equation}
where $(1i=\sqrt{-1})$ is the imaginary unit, and the DFT scaling constants and other proportionality constants are assumed to be properly handled. In short, the $p^{th}$ scattered field in $k$-space is defined by the scattering potential spectrum, $\widetilde{V}$, sampled along the $p^{th}$ offset spherical cap and properly scaled. The scattered field in the spatial domain for the $p^{th}$ illumination is thus given by,
\begin{equation}
\label{forward}
u[\cdot,\cdot,p]=\mathcal{F}_2^{-1}(\widetilde{u}[\cdot,\cdot,p]A[\cdot,\cdot]exp(1i\phi_{pupil}[\cdot,\cdot]))
\end{equation}
where $\mathcal{F}_2^{-1}$ is the 2D inverse DFT, $\phi_{pupil}[\cdot,\cdot]$ is an unknown pupil phase function that accounts for aberrations in the imaging lens (to be optimized later), and $A[\cdot,\cdot]$ is a circ function with radius $kN\!A$ that specifies the imaging lens bandpass.

Eq. \ref{forward} represents the scattered field from the sample at the microscope's image plane. As we are primarily concerned with standard microscopes that only record  total intensity, we must also consider the unscattered field, which we model as a plane wave:
\begin{equation}
\label{unscattered}
u_{back}[\cdot,\cdot,p]=u_0[p]exp(1i(k_x^{ill}[p]x[\cdot,\cdot]+k_y^{ill}[p]y[\cdot,\cdot]))
\end{equation}
where $u_0[p]$ is the optimizable amplitude of the $p^{th}$ illumination wavevector and $x[\cdot,\cdot]$ and $y[\cdot,\cdot]$ are 2D meshgrids sampling the lateral plane, consistent with the lateral spatial sampling of the scattered field (Eq. \ref{forward}). The third factor of Eq. \ref{unscattered} accounts for the direction of the $p^{th}$ illumination. Note that we do not include an optimizable phase of the unscattered field because we desire the relative phase between the scattered and unscattered fields; furthermore, any constant phase shift common to both fields will drop when we take the absolute value below.

\begin{figure}[t!]
\begin{center}
\includegraphics[width=0.6\columnwidth]{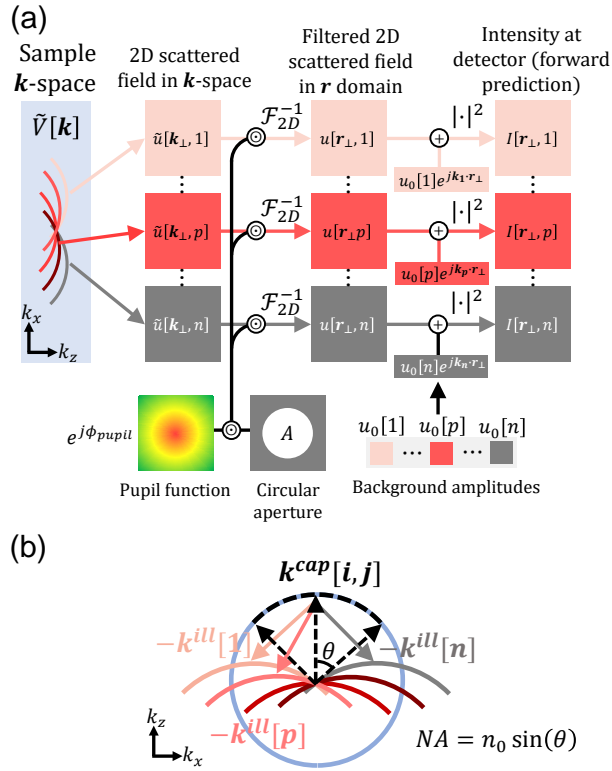}
\caption{{(a) Graphical representation of the forward model under the first Born approximation. (b) $k$-space vectorial picture, depicting the interaction between $\mathbf{k^{cap}}$ and $\mathbf{k^{ill}}$ in the Fourier diffraction theorem.
{\label{forward_appendix}}%
}}
\end{center}
\end{figure}

Finally, we arrive at the forward prediction of the image formed from the $p^{th}$ illumination LED at the image sensor under the first Born approximation:
\begin{equation}
\label{born}
I_{pred}^{Born}[\cdot,\cdot,p]=|u_{back}[\cdot,\cdot,p]+u[\cdot,\cdot,p]|^2.
\end{equation}
Fig. \ref{forward_appendix} graphically summarizes this forward model. Alternatively, under the first Rytov approximation,
\begin{equation}
\label{rytov}
I_{pred}^{Rytov}[\cdot,\cdot,p]=|u_{back}[\cdot,\cdot,p]exp(u[\cdot,\cdot,p]/u_{back}[\cdot,\cdot,p])|^2.
\end{equation}

\subsection*{Multi-slice model}
\label{multislice}
In principle, any valid forward scattering model can be used with our DP-DT procedure (replacing the ``forward model'' box in Fig. \ref{forward_figure}). Here we describe another popular scattering model, known variously as the multi-slice (MS) approximation and the beam propagation method (BPM), whereby the sample RI is parameterized as multiple thin discrete layers, within which the thin-sample approximation is assumed to apply. An incident field then propagates layer by layer, with the field emerging from the other side serving as the forward prediction. Note that this model can account for multiple scattering, but only in the forward direction.

In particular, let $\delta n_{obj}[\cdot,\cdot,\cdot]$ be a complex-valued tensor representing the sample RI deviation from the background medium RI, $n_0$. Again, the square brackets denote discrete indexing, where the dimensions correspond to the $x$, $y$, and $z$ dimensions. Let $\delta z$ be the axial sampling over a total axial sample thickness of $\Delta z$. Then, the sample can be modeled as a stack of $\Delta z/\delta z$ thin slices, separated by $\delta z$ and with phase 
\begin{equation}
\theta_{obj}[\cdot,\cdot,r]=k\delta n_{obj}[\cdot,\cdot,r]\delta z
\end{equation}
for slices $r=0, 1, ..., \Delta z/\delta z-1$. Then, given a field incident on the $r^{th}$ slice from the $p^{th}$ LED, $u_\mathbf{k_p}[\cdot,\cdot,r]$, the field exiting that slice follows the recursive relationship:
\begin{equation}
\label{base_case}
   u_\mathbf{k_p}[\cdot,\cdot,0]=u_{back}[\cdot,\cdot,p]
\end{equation}
\begin{equation}
    u_\mathbf{k_p}[\cdot,\cdot,r+1]=\mathcal{F}_2^{-1}\left(\mathcal{F}_2(u_\mathbf{k_p}[\cdot,\cdot,r])\mathcal{D}(\delta z)[\cdot,\cdot]\right)exp(1i\theta_{obj}[\cdot,\cdot,r])
\end{equation}
where $u_{back}[\cdot,\cdot,p]$ is given by Eq. \ref{unscattered}, and
\begin{equation}
    \mathcal{D}(z)[\cdot,\cdot]=exp\left(-1i\frac{(k_x^2+k_y^2)z}{kn_0+\sqrt{k^2n_0^2-k_x^2-k_y^2}}\right)
\end{equation}
is the Fresnel diffraction kernel over a distance $z$. In other words, the output field is given by the input field convolved with the diffraction kernel, followed by a phase shift due to the sample. 

We then propagate the output field to the axial position of in-focus image plane in sample space, which is then imaged by the imaging optics after the sample to the detector plane:
\begin{equation}
\begin{aligned}
    &u^{det}_\mathbf{k_p}[\cdot,\cdot]=\mathcal{F}_2^{-1}(\mathcal{F}_2(u_\mathbf{k_p}[\cdot,\cdot,\Delta z/\delta z-1]) \\
    &\times\mathcal{D}(\Delta z_f-\Delta z/2)[\cdot,\cdot]A[\cdot,\cdot]exp(1i\phi_{pupil}[\cdot,\cdot]))
\end{aligned}
\end{equation}
where $\Delta z_f$ is the position of the focus relative to the center of the sample, and $A[\cdot,\cdot]$ and $\phi_{pupil}[\cdot,\cdot]$ are the imaging optics' pupil amplitude and phase, as defined in the first Born and Rytov cases.

To avoid the edge effects that may arise from the implied circular convolutions when using DFTs, we apodize the input fields in Eq. \ref{base_case} with a Gaussian envelope
\begin{equation}
    u_\mathbf{k_p}^{apod}[\cdot,\cdot,0]=u_{back}[\cdot,\cdot,p]exp\left(-\frac{(x-x_p)^2+(y-y_p)^2}{2\sigma^2}\right)
\end{equation}
\begin{equation}
    x_p=(\Delta z_f+\Delta z/2)\frac{k_x^{ill}[p]}{k_z^{ill}[p]}
\end{equation}
\begin{equation}
    y_p=(\Delta z_f+\Delta z/2)\frac{k_y^{ill}[p]}{k_z^{ill}[p]}
\end{equation}
where the position of the Gaussian window is chosen such that after propagation to the camera conjugate plane, the resulting Gaussian-windowed field is centered within the field of view. We used a $\sigma$ value of 0.4 of the width of the FOV.

Thus, using these apodized input fields, the MS forward prediction is
\begin{equation}
    I_{pred}^{MS}[\cdot,\cdot,p]=|u^{det}_\mathbf{k_p}[\cdot,\cdot]|^2.
\end{equation}
Furthermore, the intensity data is also apodized with a centered Gaussian window of the same width to match the forward prediction:
\begin{equation}
    I_{data}[\cdot,\cdot,p]\leftarrow I_{data}[\cdot,\cdot,p]exp\left(-\frac{(x-x_p)^2+(y-y_p)^2}{2\sigma^2}\right).
\end{equation}

\subsection*{Spatial patching with the MS model}
\label{spatial_patching}
As the MS model is more computationally intensive and requires more memory than the first Born model, we found that batching along the LED dimension was insufficient to alleviate the memory constraints of the GPU. To circumvent this issue, we used what we refer to as ``spatial patching'', whereby at each iteration a uniformly random spatial patch is selected from the reconstruction (or the random noise input of DIP) along the $xy$ plane, as well as the spatially corresponding patch in the data, and the loss is computed only over that patch. In the case of DIP, as the same network is used for all patches, the spatial locations are encoded by the fixed random input to the CNN. After optimization, to create the final reconstruction, we use a stochastic stitching algorithm whereby m = 1,000 patches are randomly chosen and reconstructed, depadded to avoid edge artifacts, and superimposed.

Spatial patching, along with regular batching along the LED dimension, allows us to reconstruct arbitrarily large fields of view with limited GPU memory. Spatial patching was especially necessary when using DIP, which adds further computational overhead due to the use of multiple layers of 3D convolutions.

\begin{figure}[t!]
    \centering
    \includegraphics[width=.6\columnwidth]{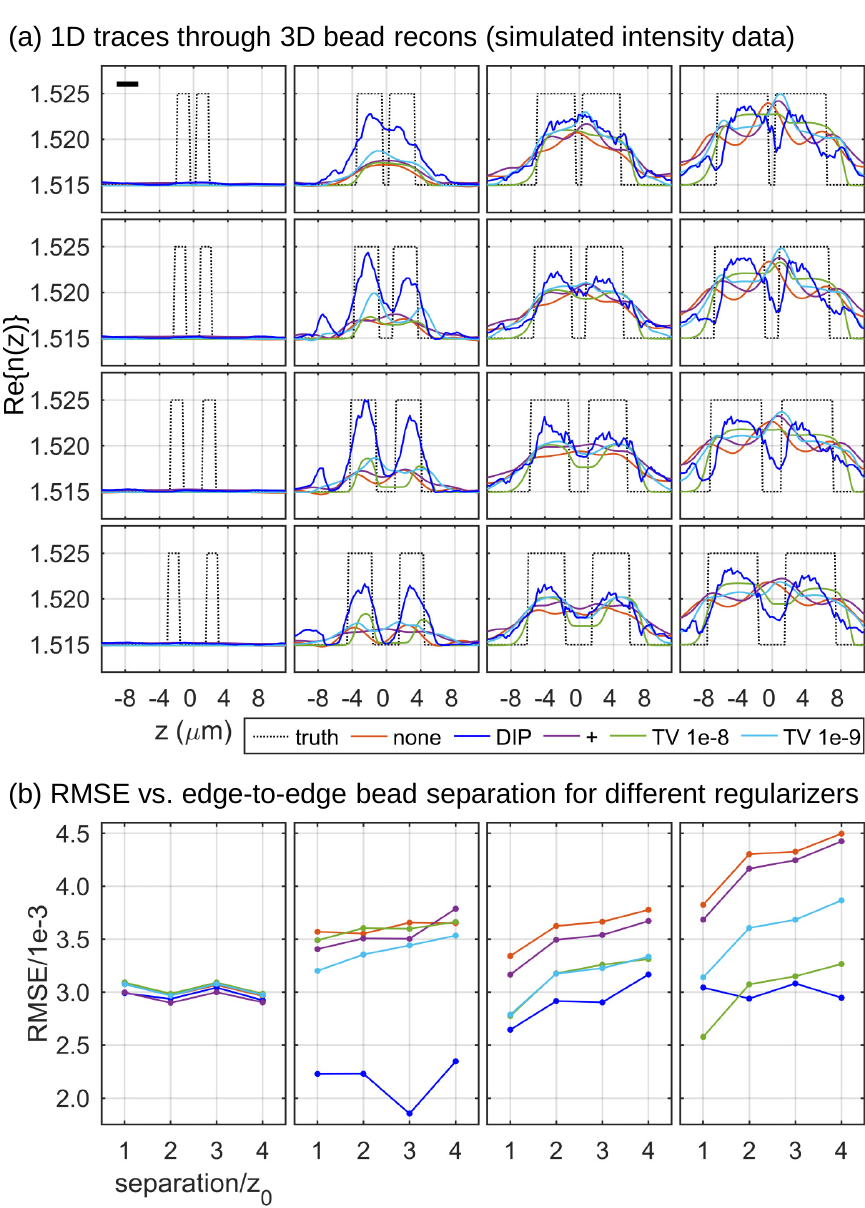}
    \caption{Comparison of the quality of 3D reconstruction from \textbf{intensity} measurements, using several regularizers for various bead sizes and separations (illumination NA = 0.4, \textbf{imaging NA = 0.1}). Figure layout is analogous to that of Fig. \ref{bead_sim}. (a) 1D traces through simulations of two beads spaced axially. Rows show different bead separations, while columns show different bead sizes. Each curve corresponds to a different regularization technique (i.e., none, DIP, positivity (+), and TV) as well as the ground truth. Scale bar corresponds to the Nyquist period. (b) The RI RMSEs from the ground truths for each regularizer. Each of the four plots corresponds to a different bead size, and each curve corresponds to a different edge-to-edge bead separation, where $z_0=$ 0.75 $\mu m$.}
    \label{bead_sim_0_1}
\end{figure}
\begin{figure}[t!]
    \centering
    \includegraphics[width=.6\columnwidth]{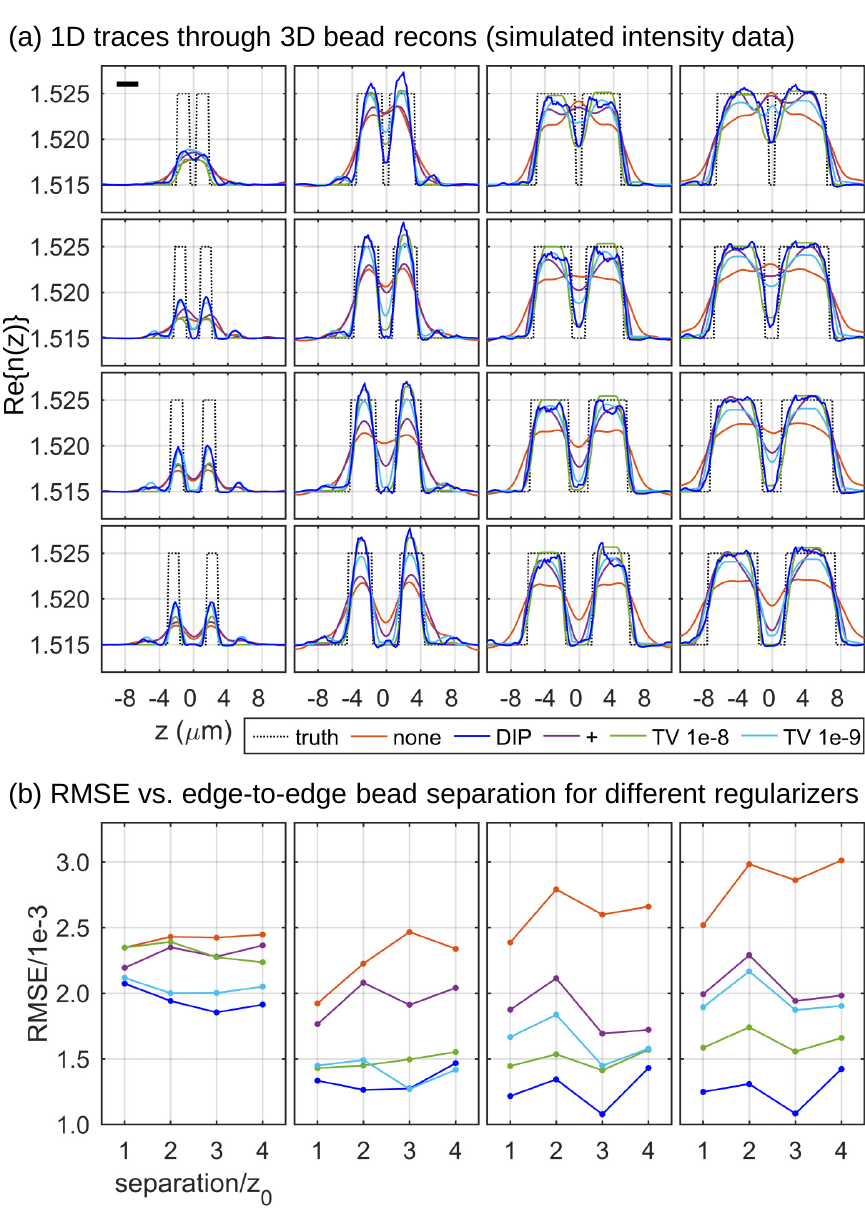}
    \caption{Comparison of the quality of 3D reconstruction from \textbf{intensity} measurements, using several regularizers for various bead sizes and separations (illumination NA = 0.4, \textbf{imaging NA = 0.3}). See caption for Fig. \ref{bead_sim_0_1}.}
    \label{bead_sim_0_3}
\end{figure}
\begin{figure}[t!]
    \centering
    \includegraphics[width=.6\columnwidth]{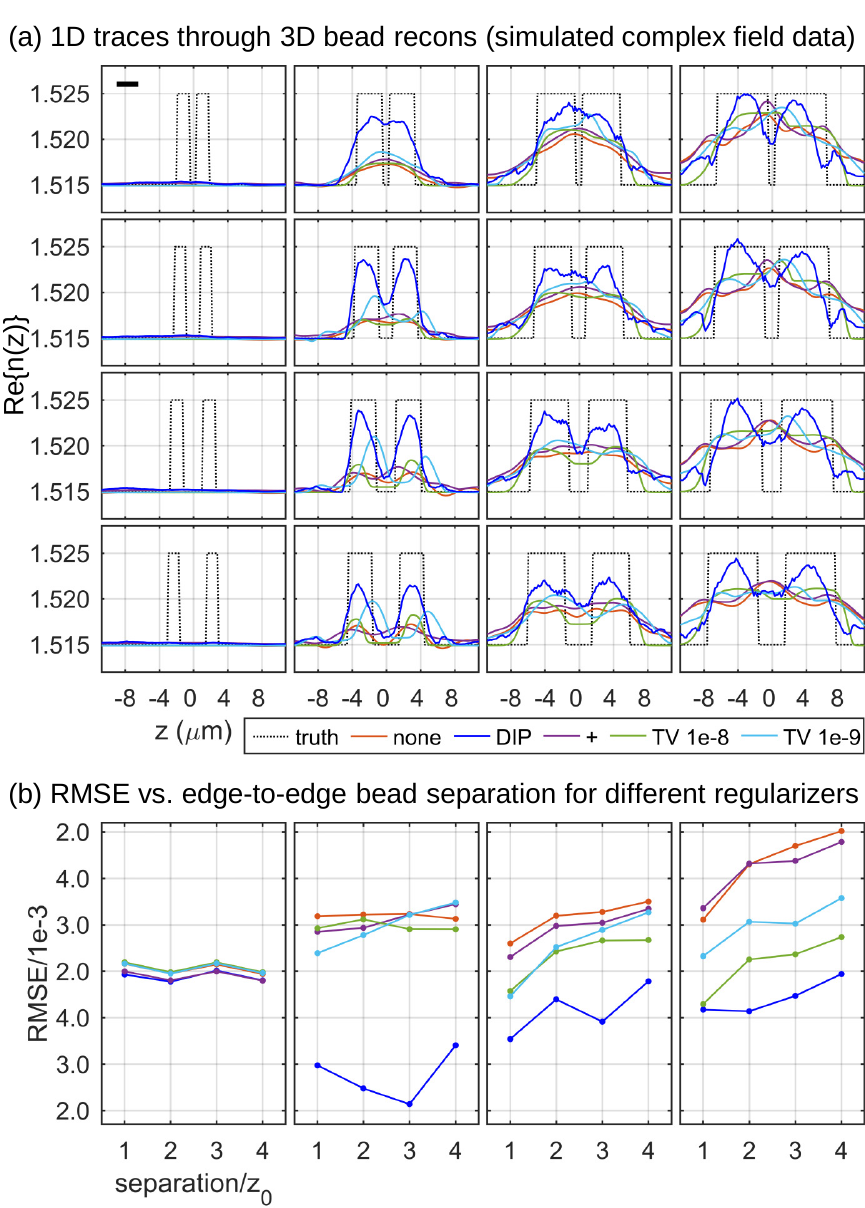}
    \caption{Comparison of the quality of 3D reconstruction from \textbf{phase-sensitive} measurements, using several regularizers for various bead sizes and separations (illumination NA = 0.4, \textbf{imaging NA = 0.1}). See caption for Fig. \ref{bead_sim_0_1}.}
    \label{bead_sim_phase_0_1}
\end{figure}
\begin{figure}[t!]
    \centering
    \includegraphics[width=.6\columnwidth]{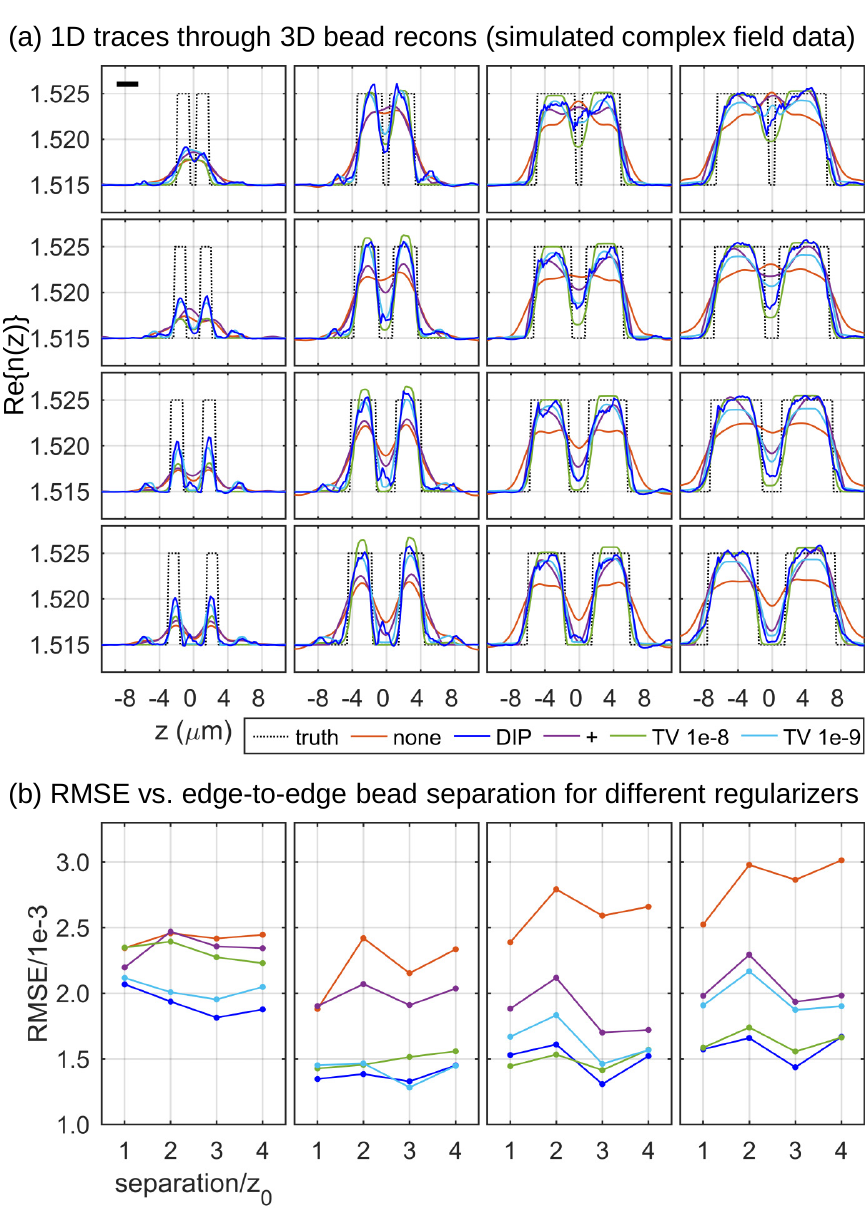}
    \caption{Comparison of the quality of 3D reconstruction from \textbf{phase-sensitive} measurements, using several regularizers for various bead sizes and separations (illumination NA = 0.4, \textbf{imaging NA = 0.3}). See caption for Fig. \ref{bead_sim_0_1}.}
    \label{bead_sim_phase_0_3}
\end{figure}
\begin{figure}[t!]
    \centering
    \includegraphics[width=.6\columnwidth]{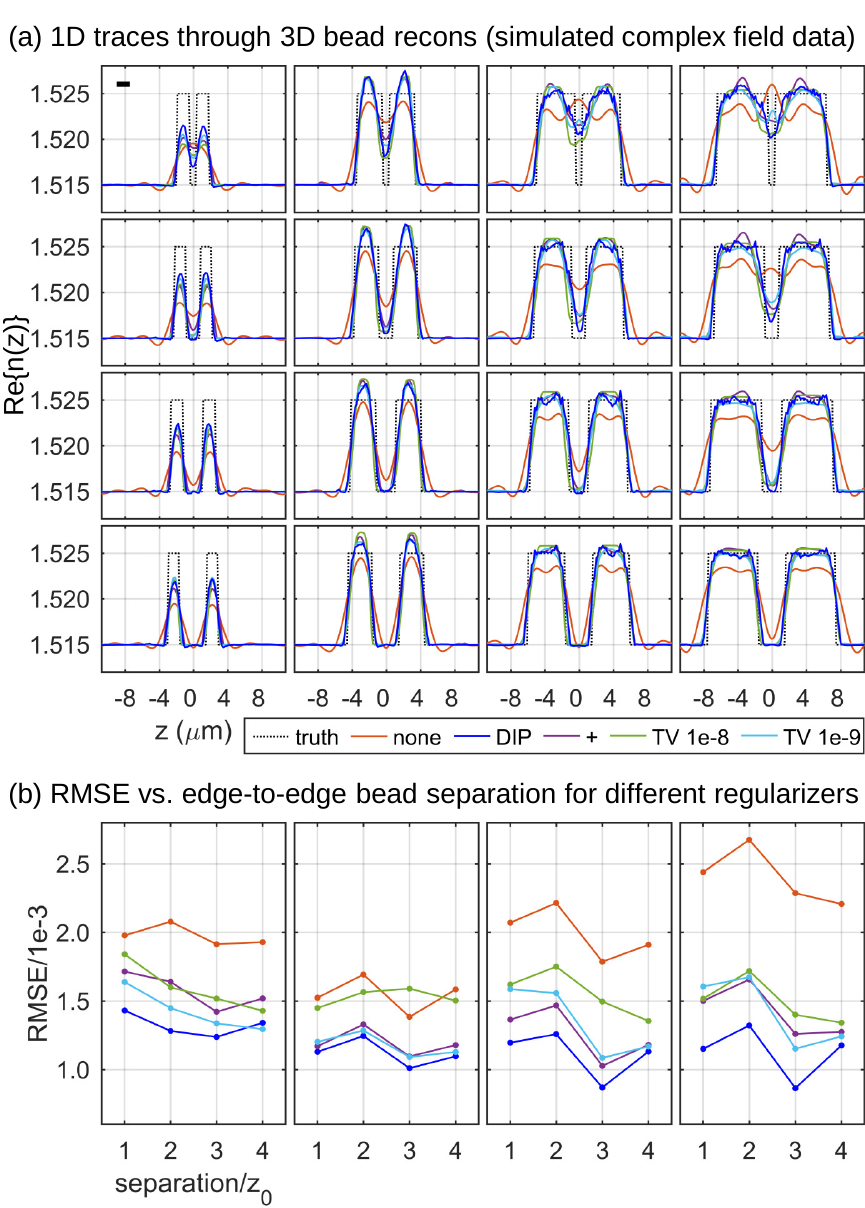}
    \caption{Comparison of the quality of 3D reconstruction from \textbf{phase-sensitive} measurements, using several regularizers for various bead sizes and separations (illumination NA = 0.4, \textbf{imaging NA = 0.5}). See caption for Fig. \ref{bead_sim_0_1}.}
    \label{bead_sim_phase_0_5}
\end{figure}

\section*{Appendix C: Additional bead simulation results}
\label{additional_results}
\subsection*{Intensity measurements (imaging NA = 0.1 and 0.3)}
Here, we provide the figures analogous to Fig. \ref{bead_sim}, which shows results for an imaging NA of 0.5, for imaging NAs of 0.1 and 0.3 (Figs. \ref{bead_sim_0_1} and \ref{bead_sim_0_3}, respectively). For these imaging NAs, DIP also outperforms other regularizers, with particularly striking results for the second column of Fig. \ref{bead_sim_0_1}.

\subsection*{Phase-sensitive measurements (imaging NA = 0.1, 0.3, and 0.5)}
\label{phase_sensitive_equations}
In the case of a phase-sensitive DT setup, we can replace the measured intensities in Eqs. \ref{born}, \ref{rytov}, and \ref{mse}, respectively, with measured optical field quantities, as follows:
\begin{equation}
f_{pred}^{Born}[\cdot,\cdot,p]=(u_{back}[\cdot,\cdot,p]+u[\cdot,\cdot,p])exp(1i\theta_{back}),
\label{born_field}
\end{equation}
\begin{equation}
f_{pred}^{Rytov}[\cdot,\cdot,p]=u_{back}exp(1i\theta_{back})[\cdot,\cdot,p]exp(u[\cdot,\cdot,p]/u_{back}[\cdot,\cdot,p]),
\label{rytov_field}
\end{equation}
\begin{equation}
E_{field}(F, \phi_{pupil},u_0)=\frac{1}{M}\sum_{i,j,p}|f_{pred}[i,j,p]-f_{data}[i,j,p]|^2,
\label{mse_field}
\end{equation}
where $\theta_{back}$ accounts for a constant phase shift due to a difference in path length between the sample and reference fields. Note that in the intensity-only case, this factor does not contribute. Thus, since this additional phase shift is a nuisance parameter, for simplicity we do not include it in our simulations.

\begin{figure}[t!]
    \centering
    \includegraphics[width=0.6\columnwidth]{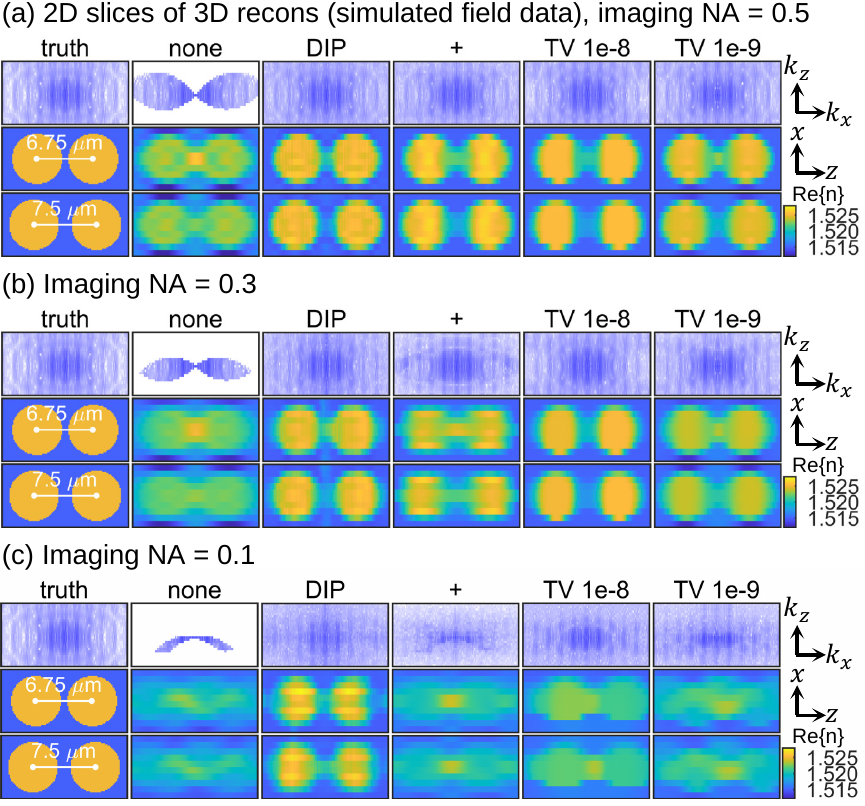}
    \caption{2D cross-sections of select simulated bead pairs, reconstructed from \textbf{phase-sensitive} measurements. Figure layout is identical to that of Fig. \ref{bead_sim2D}. (a) The first row is the through-origin $k_xk_y$ cross-sections of the scattering potential spectra of the reconstructions containing all the bead pairs, under illumination NA = 0.4, imaging NA = 0.5. The second and third rows show through-center $xz$ cross sections largest simulated bead pairs at two different separations. (b) The same information as (a), but under imaging NA = 0.3. (c) The same information as (a), but under imaging NA = 0.1.}
    \label{bead_sim2D_phase}
\end{figure}

We repeated the bead simulation experiments in Figs. \ref{bead_sim}, \ref{bead_sim_0_1} and \ref{bead_sim_0_3}, this time simulating phase-sensitive detection under similar noise conditions (see below). These results are shown in Figs. \ref{bead_sim_phase_0_1}, \ref{bead_sim_phase_0_3}, and \ref{bead_sim_phase_0_5}, which correspond to imaging NAs of 0.1, 0.3, and 0.5, respectively. Fig. \ref{bead_sim2D_phase} shows select $xz$ cross-sections. The results are very similar to those from intensity measurements, showing that DP-DT is applicable to interferometric and non-interferometric DT. Even though the data for phase-sensitive DT in principle has more information than for intensity DT, we hypothesize that the similarity of the results is due to the dense angular spacing of the illumination sources, resulting in redundancy of the $k$-space sampling. Furthermore, the $k$-space representation of relatively thin samples has redundancy in the $k_z$ direction, further increasing the redundancy of the measurements.

\subsection*{Simulating noise for the phase-sensitive measurement}
\label{noise_sim}
To simulate noise in the complex-valued datasets in order to allow fair comparison with the intensity-only scenarios, we added pixel-independent, circularly-symmetric complex Gaussian noise, such that when the intensity values are computed, the resulting quantities exhibit the same variance as predicted by photon shot noise. That is, let $I$, $E_{re}$, and $E_{im}$ be random variables representing the intensity, real part of the field, and imaginary part of the field, respectively, and let $\mu_I$, $\mu_{E, re}$, and $\mu_{E, im}$ be their respective expected value. Further, assume that all of these quantities have been properly normalized to photon counts by accounting for pixel area and integration time within a single proportionality constant, $C$; that is,
\begin{equation}
    \mu_I\leftarrow\ C^2\mu_I,\ \mu_{E, re}\leftarrow\ C\mu_{E, re},\ \mu_{E, im}\leftarrow\ C\mu_{E, im}
\end{equation}
\begin{equation}
    I\leftarrow C^2I,\ E_{re}\leftarrow CE_{re},\ E_{im}\leftarrow CE_{im}
\end{equation}
Then, we have the following:
\begin{equation}
    \mu_I=\mu_{E, re}^2+\mu_{E, im}^2
\end{equation}
\begin{equation}
    I\sim Poisson(\mu_I)
\end{equation}
\begin{equation}
    E_{re}\sim Normal(\mu_{E, re}, \sigma^2)
\end{equation}
\begin{equation}
    E_{im}\sim Normal(\mu_{E, im}, \sigma^2)
\end{equation}
We desire $\sigma$ such that
\begin{equation}
\label{condition}
    Var(E_{re}^2+E_{im}^2)=Var(I)=\mu_I
\end{equation}
where $Var(\cdot)$ denotes variance. We note that $\frac{E_{re}^2+E_{im}^2}{\sigma^2}$ follows a noncentral chi-squared distribution as follows:
\begin{equation}
    \frac{E_{re}^2+E_{im}^2}{\sigma^2} \sim \chi^2_{noncentral}(k=2, \lambda=\frac{\mu_{E, re}^2+\mu_{E, im}^2}{\sigma^2})
\end{equation}
with the following statistics:
\begin{equation}
\label{mean}
   E(E_{re}^2+E_{im}^2)=\sigma^2E\left(\frac{E_{re}^2+E_{im}^2}{\sigma^2}\right)=\mu_{E, re}^2+\mu_{E, im}^2+2\sigma^2=\mu_I+2\sigma^2
\end{equation}
\begin{equation}
\label{variance}
    Var(E_{re}^2+E_{im}^2)=\sigma^4Var\left(\frac{E_{re}^2+E_{im}^2}{\sigma^2}\right)=4\sigma^2(\sigma^2+\mu_I)
\end{equation}
where $E(\cdot)$ denotes expectation (not to be confused with field). Setting Eq. \ref{variance} equal to $\mu_I$, we arrive at
\begin{equation}
\label{result}
    \sigma=\sqrt{\frac{\sqrt{\mu_I^2+\mu_I}-\mu_I}{2}}\approx\frac{1}{2}
\end{equation}
Thus, for our field simulation experiments, we add circularly-symmetric Gaussian noise with this standard deviation to our forward predictions.

We point out a couple interesting observations about this result. According to Eq. \ref{mean}, the computed intensity is biased by $2\sigma^2$. However, perhaps more interestingly,
\begin{equation}
    \lim_{\mu_I \to \infty} \sigma = \frac{1}{2}
\end{equation}
In fact, even for $\mu_I=$ 1 photon, $\sigma\approx0.4551$, and for $\mu_I=$ 10 photons, $\sigma\approx0.4940$. Since in practice we detect far more than 1 photon, $\mu_I\gg1$, $\sigma=\frac{1}{2}$ is a very good approximation, regardless of the number of photons detected. We can see this by substituting this approximation in Eq. \ref{variance}, we have
\begin{equation}
    Var(E_{re}^2+E_{im}^2)=\mu_I+\frac{1}{4}\approx\mu_I
\end{equation}

Incidentally, since $\sigma<\frac{1}{2}$, the aforementioned bias in the expected value is in practice negligible. It's also worth noting that for large photon counts, the noncentral chi-squared distribution, like the Poisson distribution, approaches a Gaussian distribution, such that not only do the variances aysmptotically match (Eq. \ref{condition}), but also the distributions.

\bibliography{sample}

\end{document}